\documentclass[12pt]{article}

\usepackage{graphicx}
\usepackage{booktabs}
\usepackage{amsmath,amssymb}
\usepackage{epsf}

\def\ge{\mathrel{\raise.3ex\hbox{$>$\kern-.75em\lower1ex\hbox{$\sim$}}}}
\def\la{\mathrel{\raise.3ex\hbox{$<$\kern-.75em\lower1ex\hbox{$\sim$}}}}

\def\simgt{\mathrel{\raise.3ex\hbox{$>$\kern-.75em\lower1ex\hbox{$\sim$}}}}
\def\simlt{\mathrel{\raise.3ex\hbox{$<$\kern-.75em\lower1ex\hbox{$\sim$}}}}

\newcommand{\fr}[2]{\frac{#1}{#2}}

\newcommand{\nc}{\newcommand}

\nc{\gone}{\bar g_{\pi NN}^{(1)}}
\nc{\gzero}{\bar g_{\pi NN}^{(0)}}
\nc{\al}{\alpha}
\nc{\ga}{\gamma}
\nc{\de}{\delta}
\nc{\ep}{\epsilon}
\nc{\ze}{\zeta}
\nc{\et}{\eta}
\nc{\Th}{\Theta}
\nc{\ka}{\kappa}
\nc{\rh}{\rho}
\nc{\si}{\sigma}
\nc{\ta}{\tau}
\nc{\up}{\upsilon}
\nc{\ph}{\phi}
\nc{\ch}{\chi}
\nc{\ps}{\psi}
\nc{\om}{\omega}
\nc{\Ga}{\Gamma}
\nc{\De}{\Delta}
\nc{\La}{\Lambda}
\nc{\Si}{\Sigma}
\nc{\Up}{\Upsilon}
\nc{\Ph}{\Phi}
\nc{\Ps}{\Psi}
\nc{\Om}{\Omega}
\nc{\ptl}{\partial}
\nc{\del}{\nabla}
\nc{\ov}{\overline}
\nc{\newcaption}[1]{\centerline{\parbox{15cm}{\caption{#1}}}}

\nc{\hef}{\ensuremath{{}^4\mathrm{He}}}
\nc{\het}{\ensuremath{{}^3\mathrm{He}}}
\nc{\lisx}{\ensuremath{{}^6\mathrm{Li}}}
\nc{\lisv}{\ensuremath{{}^7\mathrm{Li}}}
\nc{\bes}{\ensuremath{{}^7\mathrm{Be}}}
\nc{\beet}{\ensuremath{{}^8\mathrm{Be}}}
\nc{\ben}{\ensuremath{{}^9\mathrm{Be}}}
\nc{\dm}{{\rm D}}
\nc{\hefm}{{\rm ^4He}}
\nc{\hetm}{{\rm ^3He}}
\nc{\lisxm}{{\rm ^6Li}}
\nc{\lisvm}{{\rm ^7Li}}
\nc{\besm}{{\rm ^7Be}}
\nc{\beetm}{{\rm ^8Be}}
\nc{\benm}{{\rm ^9Be}}
\nc{\bs}{(N$X^-$)}
\nc{\xm}{\ensuremath{X^-}}
\nc{\xp}{$X^+$}
\nc{\xz}{$X^0$}
\nc{\bex}{(\bes\xm)}
\nc{\bexm}{(\besm X^-)}
\nc{\px}{($p$\xm)}

\nc{\Trit}{\ensuremath{\mathrm{T}}}
\nc{\Deut}{\ensuremath{\mathrm{D}}}
\nc{\Ox}{\ensuremath{\mathrm{O}}}
\nc{\Fe}{\ensuremath{\mathrm{Fe}}}
\nc{\Hyd}{\ensuremath{\mathrm{H}}}
\nc{\Be}{\ensuremath{\mathrm{Be}}}
\nc{\tauX}{\ensuremath{\tau_{X^-}}}
\nc{\YX}{\ensuremath{Y_{X^-}}}
\nc{\YXdec}{\ensuremath{Y^{\mathrm{dec}}_{X^-}}}
\nc{\Ysldec}{\ensuremath{Y^{\mathrm{dec}}_{{\widetilde{l}_1}}}}
\nc{\Ysldecm}{\ensuremath{Y^{\mathrm{dec}}_{{\widetilde{l}_1^-}}}}

\newcommand{\fm}{\mathrm{fm}}
\newcommand{\keV}{\mathrm{keV}}
\newcommand{\MeV}{\mathrm{MeV}}
\newcommand{\GeV}{\mathrm{GeV}}
\newcommand{\TeV}{\mathrm{TeV}}
\newcommand{\seconds}{\mathrm{s}}

\newcommand{\stau}{{\widetilde{\tau}_1}}
\newcommand{\stauR}{{\widetilde{\tau}_{\mathrm{R}}}}
\newcommand{\gravitino}{{\widetilde{G}}}
\newcommand{\sel}{{\widetilde{e}_1}}
\newcommand{\electron}{e}
\newcommand{\smu}{{\widetilde{\mu}_1}}

\newcommand{\st}{{\tilde{\tau}_1}}

\newcommand{\Bino}{{\widetilde B}}

\newcommand{\gluino}{{\widetilde g}}
\newcommand{\NLSP}{\mathrm{NLSP}}
\newcommand{\Dsev}{\mathrm{D}^{\mathrm{sev}}}
\newcommand{\Dcons}{\mathrm{D}^{\mathrm{cons}}}
\newcommand{\DsevEM}{\mathrm{D}^{\mathrm{sev}}_{\mathrm{em}}}
\newcommand{\DconsEM}{\mathrm{D}^{\mathrm{cons}}_{\mathrm{em}}}

\newcommand{\HeD}{^3\mathrm{He/D}}
\newcommand{\slepton}{{\widetilde{l}_1}}
\newcommand{\sleptonR}{{\widetilde{l}_{\mathrm{R}}}}
\newcommand{\mslepton}{m_{\slepton}}
\newcommand{\lepton}{{l}}

\newcommand{\neutralino}{{\widetilde \chi}^{0}_{1}}

\newcommand{\mgr}{m_{\widetilde{G}}}
\newcommand{\mst}{m_{\tilde{\tau}_1}}
\newcommand{\mneu}{m_{{\widetilde \chi}^{0}_{1}}}
\newcommand{\mgravitino}{m_{\widetilde{G}}}
\newcommand{\mstau}{m_{\widetilde{\tau}_1}}
\newcommand{\MPl}{\mathrm{M}_{\mathrm{P}}}
\newcommand{\taustau}{\tau_{\widetilde{\tau}_1}}
\newcommand{\TRmax}{T_{\mathrm{R}}^{\max}}
\newcommand{\TR}{T_{\mathrm{R}}}

\newcommand{\monetwo}{m_{1/2}}

\newcommand{\NTP}{\mathrm{NTP}}

\newcommand{\CDM}{\mathrm{dm}}

\newcommand{\EM}{\mathrm{em}}
\newcommand{\HAD}{\mathrm{had}}

\newcommand{\OmegaDM}{\Omega_{\mathrm{DM}}}
\newcommand{\LiHprim}{^6\mathrm{Li/H}|_{\mathrm{p}}}

\newcommand{\Mpc}{\mathrm{Mpc}}
\newcommand{\km}{\mathrm{km}}
\newcommand{\Omegatp}{\Omega_{\widetilde{G}}^{\mathrm{TP}}}
\newcommand{\Omegantp}{\Omega_{\widetilde{G}}^{\mathrm{NTP}}}

\newcommand{\champ}{X^{\! -}}
\newcommand{\Ebind}{E_{\mathrm{b}}}
\newcommand{\Trec}{T_{\mathrm{r}}}
\newcommand{\aBohr}{a_{\mathrm{B}}}

\newcommand{\Rcone}{R_{\mathrm{c1}}}
\newcommand{\Rctwo}{R_{\mathrm{c2}}}
\newcommand{\Rcriti}{R_{\mathrm{c}i}}

\newcommand{\Vmax}{V_{\mathrm{max}}}
\newcommand{\Evarbind}{E_{\mathrm{b}}^{\mathrm{var}}}

\begin{document}

\begin{titlepage}
\rightline{UVIC--TH--08/13}
\rightline{MPP--2008--63}
\rightline{arXiv:0807.4287}
\begin{center}

\vspace{0.5cm}

\large {\bf Constraints on Supersymmetric Models from Catalytic Primordial Nucleosynthesis of Beryllium}
\vspace*{5mm}
\normalsize

{\bf Maxim Pospelov$^{\,(a,b)}$, Josef Pradler$^{\,(c)}$  and Frank D.~Steffen$^{\,(c)}$}

\smallskip
\medskip

{a. \it Department of Physics and Astronomy,  University of Victoria,
Victoria, BC, V8P 1A1 Canada}

{b. \it Perimeter Institute for Theoretical Physics, Waterloo,
Ontario N2J 2W9, Canada}

{c. \it Max Planck Institute of Physics,  
F\"ohringer Ring 6,
D-80805 Munich, Germany}

\smallskip
\end{center}
\vskip0.6in

\begin{abstract}
  
  The catalysis of nuclear reactions by negatively charged relics
  leads to increased outputs of primordial \lisx\ and \ben. In
  combination with observational constraints on the primordial
  fractions of \lisx\ and \ben, this imposes strong restrictions on
  the primordial abundance and the lifetime of charged relics.
  We analyze the constraints from the catalysis of \ben\ on
  supersymmetric models in which the gravitino is the lightest
  supersymmetric particle and a charged slepton---such as the lighter
  stau---the next-to-lightest supersymmetric particle (NLSP).
  Barring the special cases in which the primordial fraction of the
  slepton NLSP is significantly depleted, we find that the \ben\ data
  require a slepton NLSP lifetime of less than $6\times
  10^3\,\seconds$.
  We also address the issue of the catalytic destruction of \lisx\ and
  \ben\ by late forming bound states of protons with negatively
  charged relics finding that it does not lead to any significant
  modification of the limit on the slepton lifetime.

\end{abstract}
\vspace*{2mm}

\end{titlepage}

\section{Introduction}

Physics of the early Universe keeps proving to be an invaluable tool
for probing particle physics models and, in particular, models of New
Physics beyond the Standard Model.  One of the most important
viability checks of these models results from the epoch of Big Bang
Nucleosynthesis (BBN), i.e., from the early Universe at temperatures
of $T\simlt 1$~MeV. The combination of Standard Model physics, general
relativity, and the experimental determination of the baryon-to-photon
ratio from the anisotropies of the cosmic microwave
background~\cite{WMAPplusX} forms the framework for the standard BBN
(SBBN) which makes firm predictions for the primordial abundances of
light elements such as deuterium, helium, and lithium.  The comparison
of SBBN predictions with observationally determined primordial
fractions of these elements provides an important consistency check of
standard cosmology and serves as a remarkably powerful discriminator
among models of New Physics~\cite{SBBN}. The three most generic ways
in which New Physics can affect the outcome of BBN are the change in
the timing of the reactions caused, {\em e.g.}, by new significant
contributions to the Hubble expansion rate~\cite{SBBN}, the
non-thermal processes from late annihilation and decays of heavy
particles~\cite{nonthermalone,nonthermaltwo,Sigl:1995kk,Jedamzik:1999di,Cyburt:2002uv,Jedamzik:2004er,Kawasaki:2004qu,Jedamzik:2006xz,nonthermalthree},
and the thermal catalysis of nuclear reactions caused by
electromagnetically or strongly interacting
relics~\cite{Pospelov:2006sc}.

Catalyzed Big Bang Nucleosynthesis (CBBN) was already discussed almost
twenty years ago in Ref.~\cite{DS,champtwo}. However, only in the last two
years after the appearance of Ref.~\cite{Pospelov:2006sc}, a lot of
work has been
done~\cite{Kohri:2006cn,Kaplinghat:2006qr,Cyburt:2006uv,Steffen:2006wx,Pradler:2006hh,Hamaguchi:2007mp,Bird:2007ge,Kawasaki:2007xb,Takayama:2007du,Jittoh:2007fr,Jedamzik:2007cp,Pradler:2007is,Kersten:2007ab,Pradler:2007ar,Jedamzik:2007qk,KusakabeplusX,Pospelov:2007js,Kawasaki:2008qe,Steffen:2008bt}
in order to refine various aspects of catalysis and to understand the
implications of CBBN in the framework of specific models.
The significant interest in CBBN is also fuelled  by its direct
connection with collider physics. Indeed, the relics causing catalysis
could be directly produced at the Large Hadron Collider (LHC) because
of their electromagnetic and/or strong interactions. One of the most
interesting and perhaps one of the most natural frameworks is in this
context the supersymmetric (SUSY) extension of the Standard Model.
There a spectrum with the gravitino $\gravitino$ as the lightest
supersymmetric particle (LSP) and a charged scalar lepton
$\slepton$---such as the lighter stau $\stau$---as the
next-to-lightest (NLSP) is a commonplace occurence even if one adopts
restrictive assumptions on the soft SUSY breaking
sector~\cite{Ellis:2003dn,Cerdeno:2005eu,Jedamzik:2005dh,Cyburt:2006uv,Pradler:2006hh,Pradler:2007is,Kersten:2007ab,Pradler:2007ar}.
While the gravitino LSP is a promising candidate for dark
matter~\cite{Asaka:2000zh,Bolz:2000fu,Fujii:2003nr,Feng:2004mt,Steffen:2006hw,Pradler:2006qh,Rychkov:2007uq,Steffen:2007sp},
the charged slepton NLSP can be long-lived and thus lead to
CBBN~\cite{Pospelov:2006sc,Cyburt:2006uv,Steffen:2006wx,Pradler:2006hh,Hamaguchi:2007mp,Bird:2007ge,Kawasaki:2007xb,Takayama:2007du,Jedamzik:2007cp,Pradler:2007is,Kersten:2007ab,Pradler:2007ar,Kawasaki:2008qe}.
If such a scenario is realized in nature, each Standard Model
superpartner produced at the LHC will cascade down to the long-lived
$\slepton$ NLSP. As the lightest Standard Model superpartner, the
$\slepton$ NLSP will then appear as a quasi-stable muon-like particle
that can escape the collider detector before decaying into the
gravitino;
cf.~\cite{Ambrosanio:2000ik,Feng:2004mt,Steffen:2006hw,Ellis:2006vu,Steffen:2007sp}
and references therein.  Thus, one would find signatures that are very
different from the excess in missing energy expected in the
alternative neutralino LSP scenarios.

The most dramatic catalytic enhancement is seen in the \lisx\ and
\ben\ production triggered by the formation of bound states of \hef\
with a (generic) negatively charged relic that we call \xm.  The
catalytic path to \lisx\ and \ben\ is shown by the following sequence
of transformations~\cite{Pospelov:2006sc,Pospelov:2007js}:
\begin{eqnarray}
\label{eq:traf-Li6}
X^- \to (\hefm X^-) \to \lisxm,
\\
\label{eq:traf-Be9}
X^- \to (\hefm X^-) \to (\beetm X^-) \to  \benm.
\end{eqnarray}
Although the (\hef\xm) system has a binding energy of about 350~keV,
its formation is delayed until $T=8$~keV by an overwhelmingly large
number of energetic photons that photo-dissociate (\hef\xm). Thus, the
(\hef\xm) bound state serves essentially as a ``bottleneck'' for
\lisx\ production, whereas the path to \ben\ goes through the ``double
bottleneck'' of (\hef\xm) and (\beet\xm).  The key for the nuclear
catalysis is an enormous enhancement of the reaction rates in the
photonless recoil reactions mediated by
\xm~\cite{Pospelov:2006sc,Pospelov:2007js}:
\begin{eqnarray}
(\hefm X^-) + {\rm D} & \to & \lisxm +X^-
\label{Eq:CBBNLiSix}
\\
(\beetm X^-) + n & \to & \benm +X^-
\label{Eq:CBBNBeNine}
\ .
\end{eqnarray}
Indeed, since the rates of these catalyzed reactions exceed the SBBN
rates for the production of \lisx\ and \ben\ by many orders of
magnitude, one finds a strong sensitivity of the efficiency of the
primordial \lisx\ and \ben\ production to the abundance of \xm\ at
relevant times.
For the case of \xm\ being a thermal relic, this abundance is governed
by the \xm\ annihilation rate and by the \xm\ lifetime $\tau_{X^-}$.
In turn, observationally inferred limits on the primordial abundances
of both \lisx\ and \ben\ will impose significant constraints on the
lifetime of \xm, its mass, and its interactions.  The limits on the
lifetime and the abundance are particularly interesting in view of the
possible catalytic solution to the so-called \lisv\ problem
\cite{Pospelov:2006sc,Bird:2007ge}, which is a persistent discrepancy
between the predicted primordial amount of \lisv\ and a factor of 2--3
lower observations of \lisv\ in the atmospheres of the metal-poor
stars.

For the gravitino LSP scenarios in which \xm\ is identified with a
negatively charged slepton NLSP, $X^-=\slepton^-$, the constraints
imposed by primordial \lisx\ catalysis have already been analyzed in
detail in a number of
publications~\cite{Pospelov:2006sc,Cyburt:2006uv,Steffen:2006wx,Pradler:2006hh,Hamaguchi:2007mp,Bird:2007ge,Kawasaki:2007xb,Pradler:2007is,Kersten:2007ab,Pradler:2007ar,Kawasaki:2008qe,Steffen:2008bt}.
Assuming a standard cosmological history that leads to a typical
thermal $\slepton$ relic abundance, the bound from \lisx\ catalysis
translates into an upper limit on the $\slepton$ lifetime of
$\tau_{\slepton} = \tau_{\champ}\simlt 5\times
10^3\,\seconds$~\cite{Pospelov:2006sc,Hamaguchi:2007mp,Kawasaki:2007xb,Pradler:2007is}.
In collider-accessible regions of the parameter space, this limit is
found to be considerably more restrictive than the BBN constraints
associated with electromagnetic/hadronic energy release from
$\slepton$
decays~\cite{Cyburt:2006uv,Steffen:2006wx,Pradler:2006hh,Kawasaki:2007xb,Pradler:2007ar,Kawasaki:2008qe}.
The $\tau_{\slepton}$ limit implies a gravitino mass $\mgr$ well below
10\% of the slepton NLSP mass $m_{\slepton}$ for $\mslepton
\simlt\mathcal{O}(1\ \TeV)$~\cite{Steffen:2006wx}.  This seems to
exclude a kinematical determination of $\mgr$~\cite{Buchmuller:2004rq}
at the next generation of particle accelerators, which might have been
feasible for $0.1\,{\mslepton}\simlt\mgr<
\mslepton$~\cite{Martyn:2006as,Hamaguchi:2006vu}. Consequently, the
$\tau_{\slepton}$ limit puts a big question mark over an idea of a
collider test of supergravity via the microscopic determination of the
Planck scale~\cite{Buchmuller:2004rq}, as well as the $\mgr$-dependent
collider test of thermal leptogenesis \cite{Pradler:2006qh}.
Moreover, a post-inflationary reheating temperature above $\TR \simeq
10^7\,\GeV$ and thereby even the viability of thermal leptogenesis
with hierarchical right-handed heavy Majorana neutrinos seems to be
disfavored by the \lisx\ constraint within the Constrained Minimal
Supersymmetric Standard Model (CMSSM) for a standard cosmological
history~\cite{Pradler:2006hh,Pradler:2007is,Pradler:2007ar}.
In the CMSSM, the splitting between $\mstau$ and $\mgr$ required to
evade the $\taustau$ limit translates also into a lower limit on the
gaugino mass parameter, which is assumed to take on a universal value
$\monetwo$ at the scale of grand unification. Indeed, for the natural
gravitino LSP mass range in gravity-mediated SUSY breaking scenarios,
the cosmologically favored region can be associated with a mass range
of the colored superparticles ({\em e.g.}, a gluino mass of
$m_{\tilde{g}}\simgt 2.5~\TeV$) for which it will be very difficult to
probe SUSY at the LHC~\cite{Pradler:2007is,Pradler:2007ar}.

In this paper we analyze the constraints imposed on SUSY models by the
catalysis of a primordial \ben\ abundance. While---based on the
results of Ref.~\cite{Pospelov:2007js}---we do not expect the
constraints from \ben\ to be considerably tighter than those from
\lisx, this analysis is warranted for a number of reasons:

\begin{enumerate} 
  
\item Observations of \lisx\ are extremely difficult because its lines
  are not resolved spectroscopically with respect to the lines of
  \lisv. The claim of a ``\lisx\ plateau'' with metallicity
  \cite{Asplund} at $\sim 10\%$ of the \lisv\ abundance is being
  challenged in the recent paper~\cite{Cayrel2007}, and a new
  re-analysis of \lisx\ data is warranted as some of the observations
  may turn out to provide only upper limits. In fact, even the value
  of an upper limit on the primoridal \lisx\ abundance is subject to
  discussions: Many papers adopt upper limits on primordial
  $\lisx/\Hyd\equiv n_{^6{\mathrm{Li}}}/n_{\mathrm{H}}$ within a range
  from $10^{-11}$ to $10^{-10}$.  Unlike \lisx, \ben\ is firmly
  detected in a significant number of stars at low metallicity, and
  its observational status is not in doubt. For the latest data on the
  \ben\ abundance in metal-poor stars, see, {\em e.g.},
  \cite{Primas:2000gc,Be9,Boesgaard:2005pf}.
  
\item \lisx\ is more fragile than \lisv\ and would burn more
  efficiently at lower temperatures. Therefore, if there is a (yet
  unconfirmed) stellar mechanism (see {\em e.g.}~\cite{Korn:2006tv})
  that resolves the lithium problem, i.e., that depletes \lisv\ by a
  factor of two or three, \lisx\ would have been depleted by an even
  larger factor. Such a stellar mechanism, however, would affect \ben\
  less than either \lisv\ or \lisx\ since both \lisv\ and \lisx\ are
  more fragile than \ben, which thereby provides a more robust bound
  on New Physics.

\item The nuclear physics rates that enter in the calculation of \ben\
  catalysis are dominated by resonances. Given the wealth of
  experimental information on the \ben\ resonances \cite{Sumiyoshi},
  this may eventually allow for very reliable calculations of the
  catalytic rates.
  
\end{enumerate}

While stating strong bounds on primordial abundances/lifetimes of \xm,
a lingering question remains: Is there an ``island'' of allowed
lifetimes around $\tau_{X^-} \sim 10^6~\seconds$?
One point emphasized in Refs.~\cite{DS,Kohri:2006cn,Jedamzik:2007cp}
is that bound states of \xm\ with protons, \px, may have a significant
impact on the primordial abundances of elements such as \lisx, \lisv,
\bes\ (and \ben).
Another point is related to \xm\ decays with very energetic decay
products and the associated possibility of an environment in which
\lisx\ may be destroyed efficiently.
The second point has already been
addressed~\cite{Cyburt:2006uv,Kawasaki:2007xb,Kawasaki:2008qe}, and it
has been found that late energy injection cannot suppress \lisx\ down
to an acceptable level if it is significantly overproduced at
8~keV~\cite{Cyburt:2006uv,Kawasaki:2007xb,Kawasaki:2008qe}
Despite its importance, the issue of possible ($p$\xm) catalysis
remained largely unresolved.  While the importance of charge exchange
reactions of \px\ with \hef\ that may reduce the abundance of ($p$\xm)
was already mentioned in the early paper~\cite{DS}, subsequent
publications either ignored this issue~\cite{Kohri:2006cn} or
generally underestimated the impact of charge exchange
reactions~\cite{Jedamzik:2007cp}.
Recognizing its importance for the whole CBBN paradigm, we revisit the
catalysis by ($p$\xm) bound states.
Indeed, in this paper, we put this issue to rest by reaching the
conclusion that less than 10\% of \lisx\ and \ben\ synthesized at 8
keV could possibly be affected by \px\ catalysis, whereas typically
suppression factors in excess of 100 are needed in order to evade the
corresponding limits.
This clarifies that the islands in the parameter region with large
abundances/large lifetimes, which were suggested to remain
viable in Ref.~\cite{Jedamzik:2007cp}, cannot exist.

This paper is organized as follows. In the next section we estimate
the charge exchange reaction rates relevant for ($p$\xm)--mediated
catalysis and consider their consequences for lithium and beryllium.
In Sect.~\ref{Sec:9BeConstraints} we analyze the bounds on the
lifetime/abundance of \xm\ imposed by observations of \ben\ in stellar
atmospheres at low metallicities. The resulting constraints on SUSY
models with the gravitino LSP and a charged slepton NLSP are given in
Sect.~\ref{Sec:SUSYImplications}. We reach our conclusions in
Sect.~\ref{Sec:Conclusions}.

\section{\boldmath($p$\xm) catalysis and charge exchange reactions}
\label{Sec:pXcatalysis}

We begin the discussion of the possibility of \px--mediated catalysis
by reminding the reader of the basic properties of the \px\ and
(\hef\xm) bound states. Table~\ref{table1} lists the corresponding
binding energies $\Ebind$, Bohr radii $\aBohr$, and recombination
temperatures $\Trec$, where $\Trec$ is understood as the temperature
below which the rate of photo-dissociation of a bound state becomes
smaller than the Hubble expansion rate.
%
\begin{table}
\caption{
\footnotesize Properties of \px\ and (\hef\xm) bound states. 
For (\hef\xm), the binding energy $\Ebind$ includes 
a finite charge radius correction \cite{Pospelov:2006sc}.
Bohr radii $\aBohr$ are quoted for idealized Bohr-type bound states. 
The given recombination temperatures $\Trec$ are understood
as the temperatures at which the corresponding photo-dissociation rate 
becomes equal to the Hubble rate.
}
\begin{center}
\begin{tabular}{cccc}
\toprule\\[-0.25cm]
bound state & ~$\Ebind$(keV)~ & ~$\aBohr$(fm)~ & 
~$\Trec$(keV)~ \\[0.1cm]
\midrule\\[-0.30cm]
(\hef\xm)&     $-347$      &  3.6        &     8\\[0.1cm] 
\px &          $-25$       &   29        &   0.6 \\[0.1cm]
\bottomrule
\end{tabular}
\end{center}
\label{table1}
\end{table}
%
As the table suggests, the
recombination of \px\ bound states becomes efficient only after the
temperature drops below 1~keV which corresponds to an age of the
Universe on the order of a few weeks or so, and we will assume in this
section that the \xm\ lifetime $\tau_{\champ}$ is large enough to
allow for this recombination to happen. Clearly, the presence of a
large number density of negatively charged particles $n_{\champ}$
during the recombination with helium leads to an
overproduction of \lisx\ and
\ben~\cite{Pospelov:2006sc,Pospelov:2007js} due to the 
nuclear catalysis at 8 keV. To be specific, we choose
$Y_{\champ}\equiv n_{\champ}/n_{\mathrm B} = 10^{-2}$ (with
$n_{\mathrm B}$ denoting the baryon number density),
$\tau_{\champ}\to\infty$, and using
our previous
results~\cite{Pospelov:2006sc,Pospelov:2007js,Pradler:2007is} 
determine the
abundances of lithium and beryllium at $T=1$~keV from CBBN production:
\begin{equation} 
\lisxm/{\rm H}|_{T=1\,\keV} \simeq 8\times 10^{-8} \ ;~~~~
\benm/{\rm H}|_{T=1\,\keV} \simeq 3\times 10^{-10} \ ,
\end{equation}
which is about three orders of magnitude above the observational
bounds; cf.\ Sect.~\ref{Sec:9BeConstraints} below.  Some of the
synthesized \lisx\ and \ben\ will be in bound states with \xm.

Below $T=1$~keV, the concentration of \px\ is controlled by a rapidly
diminishing photo-dissociation rate and by depletion through the
charge exchange reaction~\cite{DS}:
\begin{equation} 
\label{pXHe}
(pX^-)+\hefm\to (\hefm X^-) +p.
\end{equation}
This reaction may have a very large rate as its cross section is
determined by the actual size of the \px\ bound state that is of the
order of $\aBohr\simeq 30$~fm (cf.~Tab.~\ref{table1}). In fact, most of
the recombined states \px\ are immediately intercepted by the
reaction~(\ref{pXHe}) so that the resulting \px\ abundance would
remain quite small at all temperatures.

From studies of charge exchange reactions of muons on hydrogen, it is
known that the muon is captured into highly excited states that have
large orbital momenta and large principle quantum numbers. The radii
of these excited orbits of muonic hydrogen are comparable to the Bohr
radius of ordinary hydrogen.  In the case of~(\ref{pXHe}), the capture
would mainly proceed to the $n=3$ and $n=4$ levels of the (\hef\xm)
bound state.  To estimate the cross section for the reaction
(\ref{pXHe}), we employ a semiclassical approximation in which the
motion of helium is described by a classical trajectory while the
proton is treated quantum mechanically.  The large values of $n$ and
$l$ of the resulting (\hef\xm) bound states give some justification to
this treatment.  

\begin{figure}[t]
\centerline{\includegraphics[width=10cm]{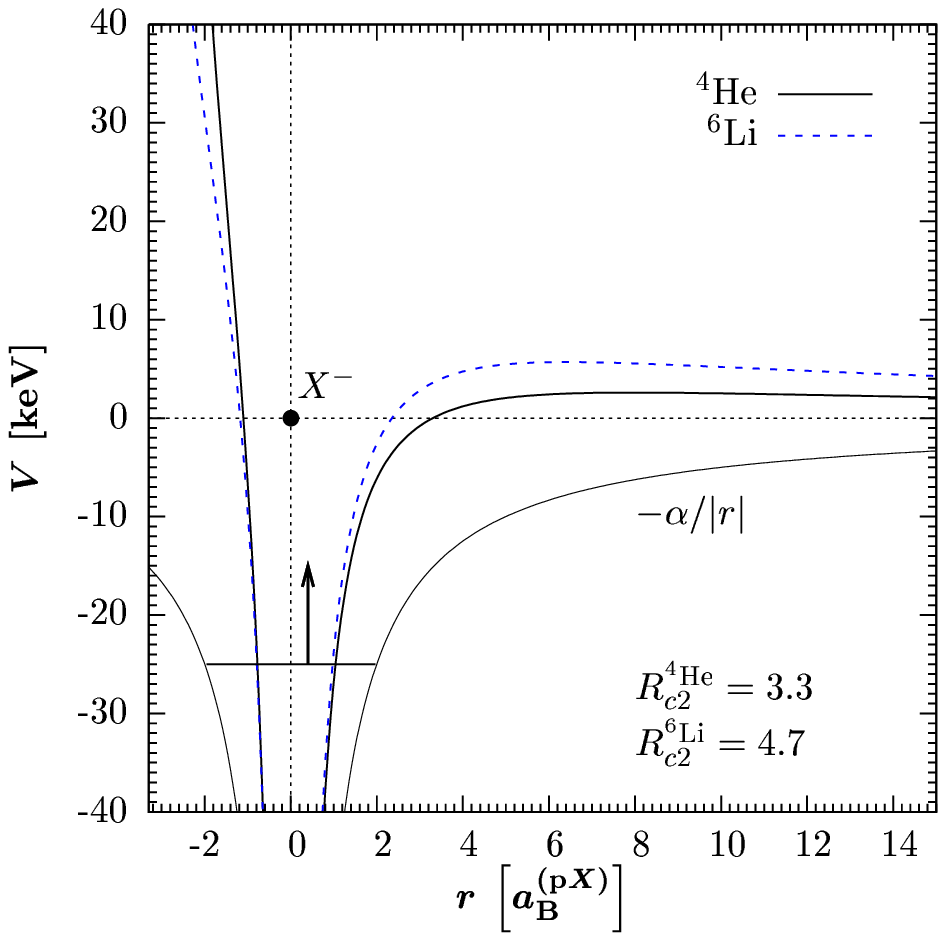}}
 \caption{
   \footnotesize Potential energy of the proton in the field of \xm\ 
   at $r=0$ and an incoming nucleus at $r=-R_{c2}$. The potential
   energy is plotted along the line connecting \xm\ with \hef\ (solid
   line) or \lisx\ (dashed line), respectively.  As the distance
   between the incoming nucleus and $X^-$ decreases, the potential
   well becomes more narrow, and the proton ground state energy level
   is pushed upward. The critical deconfinement distance
   $R_{c2}^{^4\mathrm{He},^6\mathrm{Li}}$ is defined as the distance
   at which the energy of the bound state found variationally
   using~(\ref{wf}) becomes larger than the height of the barrier
   $V_{\mathrm{max}}$ to the right of \xm.}
\label{Fig:Potential} 
\end{figure}

Calling $R$ the separation between \hef\ and \xm\ (or, more generally,
the separation between \xm\ and the incoming nucleus of charge $Z$),
we now investigate the $R$ value at which the proton loses its ability
to bind to \xm.  The one-dimensional slice of the proton potential
energy in the field of \xm\ and \hef,
\begin{equation} 
\label{V}
V({\bf r}) = -\fr{\alpha}{r} + \fr{\alpha\,Z}{|{\bf r} - {\bf R}|},
\end{equation}
is plotted in Fig.~\ref{Fig:Potential}. 
%
%
The limit of $R\to\infty$ corresponds to an unperturbed binding of the
proton to \xm\ with a binding energy of $\Ebind=-25~\keV$
(cf.~Tab.~\ref{table1}). For $Z>1$ and finite $R$, the curve has a
maximum at positive values of $r$ referred to as $\Vmax$. As the \hef\ 
nucleus comes closer, $R$ decreases. For $R$ values below some
critical distance $\Rcone$, the binding energy of the proton becomes
positive so that the tunneling of the proton to $r\to+\infty$ starts
to become viable.
For even smaller values of $R$, one can find another ``critical''
distance $\Rctwo$ at which the probability for the tunneling of the
proton becomes comparable to 1 due to the fly-by of the \hef\ nucleus.
In principle, this is not an easy quantum mechanical problem, and we
use further simplifications to overcome that.  In order to estimate
$\Rcriti$, we employ the variational calculation of the proton energy
in the potential (\ref{V}) by using the trial wave function for the
ground state,
\begin{equation} 
\label{wf}
\psi(\mu, \nu) = \exp[-(\mu-\nu)R/(2a)]\times(1+\nu R/b)^2,
\end{equation}
where $\mu$ and $\nu$ are elliptic coordinates and $a$ and $b$ the
minimization parameters. The coordinates are defined as 
$\mu=(r_1+r_2)/(2R)$ 
and 
$\nu = (r_1-r_2)/(2R)$, 
where $r_1$ and $r_2$ are the proton--nucleus and proton--\xm\ 
distances, respectively. We calculate the energy of the ground state
$\Evarbind$ as a function of the distance between \xm\ and the
incoming nucleus $R$. We determine $\Rcone$ from $\Evarbind(\Rcone)=0$
and estimate $\Rctwo$ from $\Evarbind(\Rctwo)=\Vmax$ 
which describes the situation when even a metastable bound state
simply cannot exist.
The cross section for the charge exchange reaction is then
approximated by the geometric one with the impact parameter
$\rho=\Rctwo$, $\sigma = \pi \Rctwo^2$, which essentially assumes a
deconfinement probability of 1 for $R\leq\Rctwo$. This approximation
would fail for very fast incoming nuclei. For the case considered in
this paper, however, the fly-by time is much longer than the period
with which the proton orbits within \px.
The results of our estimates are presented in Table~\ref{table2}. 
%
\begin{table}
  \caption{ \footnotesize Deconfining distances and charge exchange reaction
    cross sections on the \px\ target for incoming nuclei with different charges $Z$. 
    The $\Rcriti$ values are given in units of the \px--Bohr radius $a_{\mathrm{B}}^{(pX^-)}$ 
    and in units of fm.}
\begin{center}
\begin{tabular}{cccc}
\toprule\\[-0.25cm]
$Z$ & ~$\Rcone$ at $E=0$~ & ~$\Rctwo$ at $E=\Vmax$~ & 
~$\sigma=\pi\Rctwo^2$ ~in bn \\[0.1cm]
\midrule\\[-0.30cm]
1 &        1.4 (40 fm)  &  1.4 (40 fm)   & 51 bn    \\[0.1cm]
2 &        3.7 (107 fm) &   3.3 (95 fm) & 280 bn    \\[0.1cm]
3 &        5.8  (167 fm)&  4.7  (135 fm) &  580 bn  \\[0.1cm]
4 &        7.9 (230 fm) &   5.7 (160 fm) &  850 bn \\[0.1cm]
\bottomrule
\label{table2}  
\end{tabular}
\end{center}
\end{table}
%
As can be seen from this table, a \hef--\xm\ distance of $\sim 95~\fm$
is sufficient to release the proton from the bound state.
Consequently, our estimate points to a very large cross section of
almost $300$~bn for the charge exchange reaction~(\ref{pXHe}).
For $Z=1$ we have a cross section $\sim 2\pi\aBohr^2$, which compares
well with the results for the charge exchange cross section in the
case of muon--hydrogen scattering~\cite{pra}.

Using this cross section, we incorporate the charge exchange reactions
in the network of Boltzmann equations and calculate the residual
concentration of \px\ in a wide temperature range. In the limit of
infinite lifetimes, $\tau_{X^-}\to\infty$, we find that the abundance
of \px\ reaches its peak at around $T=0.7$ keV.  Its maximum abundance
at these temperatures can be well approximated as
\begin{equation} 
\frac{n^{\rm max}_{(pX^-)}}{n_p} 
\simeq 
4\times 10^{-7} \left(\frac{\YX}{10^{-2}}\right),
\label{pXmax}
\end{equation}
where we made the safe assumption%
\footnote{ Focusing on a minimal MSSM particle content, \YX\ is
  determined by the standard chemical decoupling of $\champ$ from the
  primordial plasma for which $\YX \la Y_{\rm ^4He}$ holds unless
  $m_{\champ}\gtrsim\mathcal{O}(4\ \TeV)$; see
  Sect.~\ref{Sec:SUSYImplications} for a detailed discussion of \YX.}
of $\YX \la Y_{\rm ^4He}$, which ensures the linear scaling in
(\ref{pXmax}).

Reference~\cite{Jedamzik:2007cp} makes the somewhat surprising
suggestion that even a tiny fraction of surviving \px\ bound states
may cause a significant reduction of the \lisx\ abundance, given the
significant uncertainty in the nuclear rates. It is easy to see,
however, that two types of processes are possible for the colliding
\lisx--\px\ system: the charge exchange reaction and the nuclear reaction,
\begin{eqnarray} 
{\rm A:}~~~~(pX^-)+\lisxm &\to& (\lisxm X^-) +p
\, ,
\label{Eq:pXLi}
\\
{\rm B:}~~~~(pX^-)+\lisxm &\to& X^- +\hefm+\hetm
\, .
\end{eqnarray}

At first, we completely ignore process A and concentrate on process B
which destroys \lisx.  To illustrate our point, we will assign the
maximal possible rate to process~B, which is given by the unitarity
bound in the $s$-channel,
\begin{equation} 
\langle \sigma v \rangle^{\rm max}_{\rm B} = \frac{\pi}{m_{\lisxm}^2}\langle v^{-1} \rangle 
=\fr{\sqrt{2\pi}}{m_\lisxm\sqrt{m_\lisxm T}} = \fr{1.4\times 10^8}{T_9^{1/2}},
\label{maximal}
\end{equation}
where the last expression is calculated in units of
$N_A^{-1}$cm$^3\,\seconds^{-1}$mol$^{-1}$, and $T_9 = T/10^9$~K.
Comparing the destruction rate to the Hubble rate at a fiducial
temperature of $T_9 = 0.008$, we find that the former is much smaller,
\begin{equation}
\label{comparetoH}
\left.\fr{\langle \sigma v \rangle^{\rm max}_{\rm B}n^{\rm max}_{(pX^-)}}{H}\right|_{T_9=0.008} 
\simeq 0.03 \ll 1
\end{equation}
for $\YX=0.01$.
This tells us that at most 3\% of the \lisx\ synthesized at 8 keV
could potentially be affected by \px\ states via process B, and
therefore the whole issue of nuclear uncertainties is irrelevant given
the strength of the charge exchange reactions.

Even though we find that process B makes no impact on \lisx, it is
still interesting to examine whether (\ref{maximal}) corresponds to a
realistic rate.  As can be seen from Fig.~\ref{Fig:Potential}, for
reaction B to occur, the proton has to tunnel to the left through a
distance of at least 100~fm inside the Coulomb barrier of \lisx\ in
order to trigger its decay to two helium nuclei. Notice that the field
of \xm\ does not facilitate this tunneling in any way. Effectively,
this is the same distance for tunneling that a {\em free} $\sim
50~\keV$ energy proton would have to overcome in a $p+\lisxm\to
\hefm+\hetm$ reaction.  Therefore, one expects an exponential
suppression of the corresponding probability by
$\exp(-\sqrt{E_G/50~{\rm keV}})\ll 1$, where $E_G$ is the Gamow energy.  At the same time, the
deconfining rate is 100\% as long as the impact parameter is less or
equal to $\Rctwo$, leading to the inevitable conclusion that the rate
of the charge exchange reaction~A greatly exceeds that of the nuclear
process~B,
\begin{equation}
\label{B/A}
\fr{\langle \sigma v \rangle_{\rm B}}{\langle \sigma v \rangle_{\rm A}}
\ll 1.
\end{equation}
Returning to the calculation of Ref.~\cite{Jedamzik:2007cp} that gives
a very large estimate for process B, we believe that this estimate is
probably an artefact of assuming a ``frozen'' profile for the proton
wave function. In contrast, in the correct approach, the proton wave
function is easily polarized and deconfined by the incoming heavier
nucleus. Further doubts in the validity of the estimates
in~\cite{Jedamzik:2007cp} are cast by the answer for the cross section
at extremely small energies.  For example, it follows
from~\cite{Jedamzik:2007cp} that $\sigma_{\rm B}$ has almost an atomic
size cross section, whereas this is a nuclear reaction between objects
of nuclear size.  The key difference between our treatment of
\px--induced catalysis and that of Ref.~\cite{Jedamzik:2007cp} is the
significant underestimation of the strength of the charge exchange
reactions in the latter work, which in turn leads to overestimates in
expressions analogous to our Eqs.~(\ref{pXmax})
and~(\ref{comparetoH}).

To conclude this section, we consider an interesting way of having an
impact of \px\ on \lisx. Indeed, a successive chain of charge exchange
reactions can lead to molecular states that are finally destroyed in
nuclear reactions with protons:
\begin{eqnarray}
(pX^-)+\lisxm \to (\lisxm X^-) +p\\
(pX^-)+(\lisxm X^-)\to (\lisxm X^-_2) +p\\
(pX^-)+(\lisxm X^-_2)\to (\lisxm X^-_3) +p\\
(\lisxm X^-_3) + p \to \besm +3X^- ~{\rm or} ~ \hefm+\hetm +3X^-.
\end{eqnarray}
In the last step of this chain, $(\lisxm X^-_3)$ ``ammonium'' has a
chance for a nuclear interaction with protons or helium unsuppressed
by a residual Coulomb barrier since $(\lisxm X^-_3)$ is a very compact
object. A similar chain exists for \bes\ and \ben\, where the sequence
of the charge exchange reactions can proceed until Be--$X^-$
``methane,'' $(\benm X^-_4)$. It is important to note that the
efficiency of this chain reaction depends very sensitively on the
concentration of the \px\ bound states and on the mass of the \xm\
particle. The latter enters through the average relative velocity of
two heavy objects, {\em e.g.}, $(pX^-)$ and $(\lisxm X^-)$, which in
turn scales as $m_{X^-}^{-1/2}$.  Therefore, in the limit of an
infinitely heavy $X^-$ and with our treatment of the charge exchange
reactions, the chain will be cut off right at the first step,
terminating at (\lisx\xm).  For weak scale relics, the suppression of
the average velocity of \xm-containing bound states relative to the velocity of light nuclei
is from one to two orders of magnitude.

\begin{figure}[t]
\centerline{\includegraphics[width=11cm]{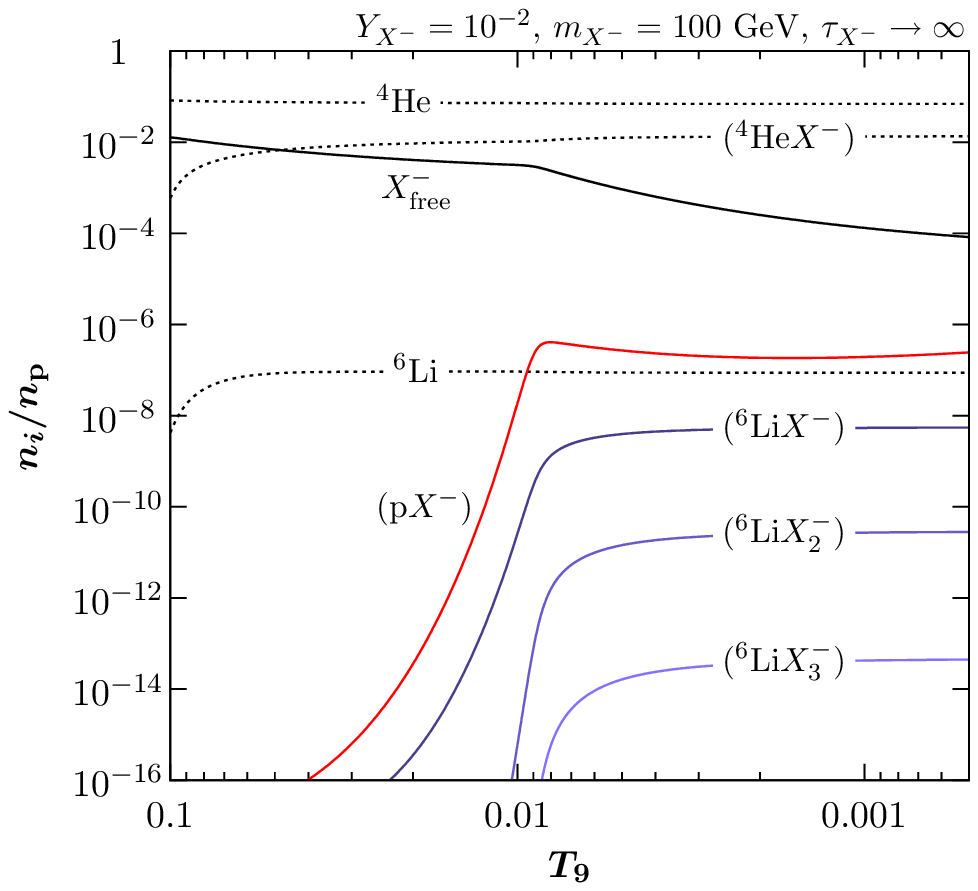}}
 \caption{
   \footnotesize Evolution of primordial abundances as a function of
   time (or temperature $T_9$) from the input $Y_{X^-}=0.01$,
   $m_{X^-}=100~\GeV$, and $\tau_{X^-}\to\infty$. The solid (red in
   the web version) line of the \px\ abundance reaches its maximum of
   $\sim 4\times 10^{-7}$ at $T_9\simeq 0.008$.  Around the same
   temperatures, the yield of unbounded $X^-$, $Y_{X^-}^{\rm free}$,
   starts to decline more rapidly since it is removed by the
   recombination with $p$ followed by the charge exchange reaction on
   \hef.}
\label{f2} 
\end{figure}

Using the charge exchange rates estimated in this section, we run the
set of Boltzmann equations to determine the residual concentrations of
\px\ and of the molecular bound states of \lisx\ with \xm. The results
are plotted in Fig.~\ref{f2}. 
%
%
As one can see, an initial concentration of \xm\ per nucleon of $\YX =
10^{-2}$ results in a \px\ abundance that never exceeds the
maximum~(\ref{pXmax}), leading to a progressively diminishing number
of molecular states.

There is another plausible mechanism for the destruction of (\lisx\xm)
that is related to the recoil of \lisx\ freed in the decay of \xm.
Typical kinetic energies of \lisx\ after the decay are comparable to
that on the orbit $\sim 700~\keV$. In the center of mass frame with
plasma protons, this corresponds to energies of $\sim 100~\keV$. At
such a center of mass energy, some of the \lisx\ nuclei released in
the decays of the bound states will be destroyed.  However, given that
only a small fraction of \lisx\ is locked in bound states with \xm and
that the destruction rate is smaller than the thermalization rate, it
is safe to conclude that also this mechanism cannot lead to large
overall depletion factors for \lisx. The same argument applies to
\ben.

To conclude this section, neither lithium nor beryllium synthesized in
CBBN processes at $8~\keV$ would be affected in any significant way by
the subsequent generation of \px\ bound states.
Thus, the part of the parameter space with a typical freeze-out \xm\
abundance and a long \xm\ lifetime is confidently ruled out, which is
shown in more detail in the following sections.

\section{\boldmath\ben\ constraints on the \boldmath\xm\ lifetime and abundance}
\label{Sec:9BeConstraints}

%
In order to constrain the (\tauX,\,\YX) parameter space from the
catalytic path (\ref{eq:traf-Be9}) to \ben~\cite{Pospelov:2007js}, we
need to set an \textit{upper limit} on its primordial abundance from existing observations.
It is generally accepted that the galactic evolution of the abundances
of Be, along with Li and B, are dominated by cosmic-ray
nucleosynthesis.  While Be is burned rapidly in stellar centers, it is
produced in cosmic rays by the spallation reactions of fast protons
and $\alpha$ particles hitting ambient CNO nuclei~\cite{ReevesGCRX}.
As a consequence, the abundances of Be and O are linked, leading to a
secondary scaling, Be $\propto\, \Ox^2$~\cite{VFetal}. On the other
hand, inverse spallation reactions of CNO nuclei, both produced and
accelerated in supernovae, will give a Be yield that is essentially
independent of the metallicity of the interstellar medium.  Such
primary processes, leading to Be $\propto\, \Ox$, are expected to play
a major role during the early galactic epochs~\cite{Fields:1999ib}.

The produced Be is subsequently supplemented in the outer layers of
stars. Thus, old stars which are far from the galactic center (and
thereby less affected by the galactic chemical evolution) bear the
potential to encode any pre-galactic origin of Be. Indeed, Be has been
observed in a number of Population II halo stars at very low
metallicities $[\Fe/\Hyd] \lesssim -2.5$; $[A]\equiv\log_{10}{A}+12$.
Particulary noteworthy is the detection in the star G~64--12 at
$[\Fe/\Hyd] \simeq -3.3$~\cite{Primas:2000gc}. The star's high Be
value of $\log_{10}(\Be/\Hyd) \simeq -13.05$ might suggest a possible
flattening in the Be trend during the early evolutionary phases of our
galaxy~\cite{Primas:2000gc}. Whether this really points to a
\textit{primordial} plateau or whether this indicates a Be dispersion
at lowest metallicities~\cite{Boesgaard:2005pf} is not clear at
present.

\begin{figure}[t]
\centerline{\includegraphics[width=0.49\textwidth]{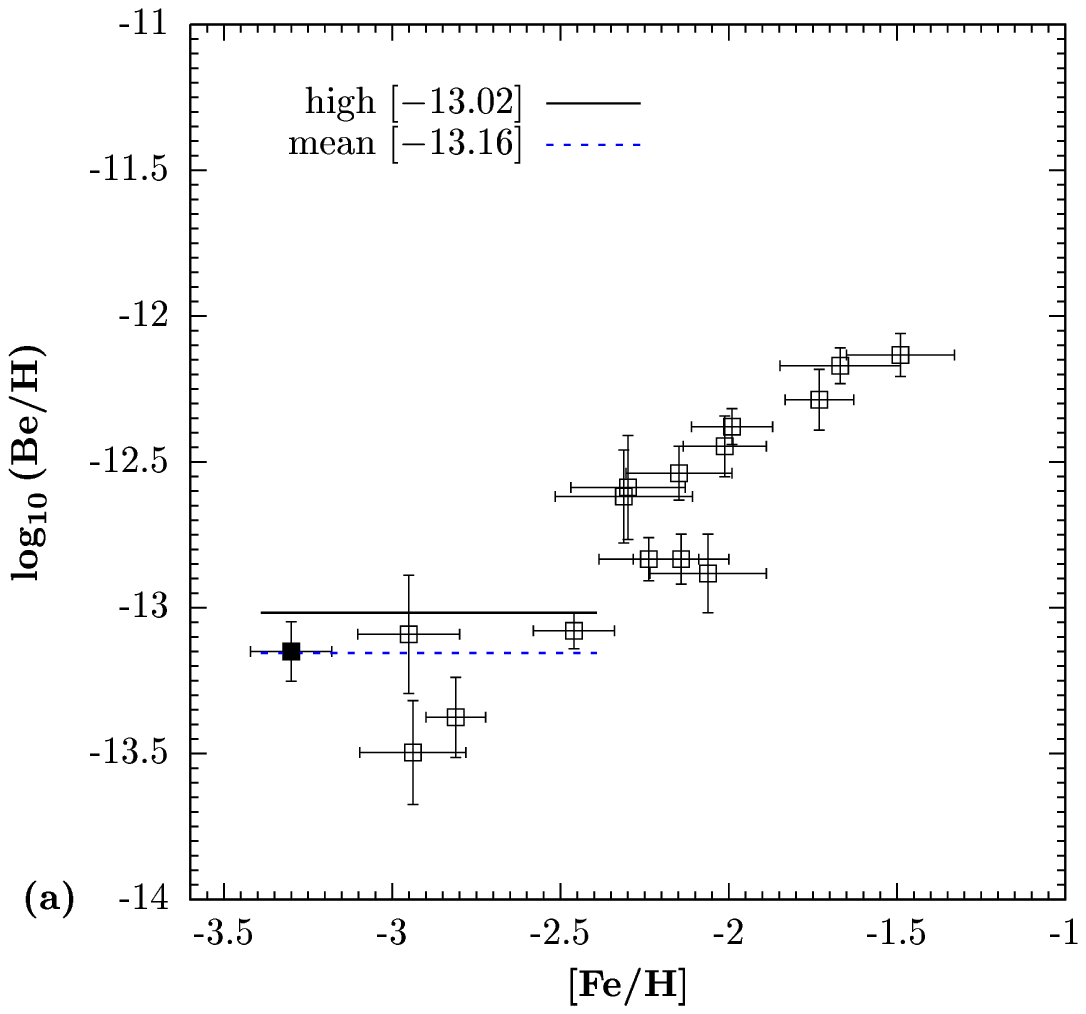}%
\hskip 0.3cm
\includegraphics[width=0.49\textwidth]{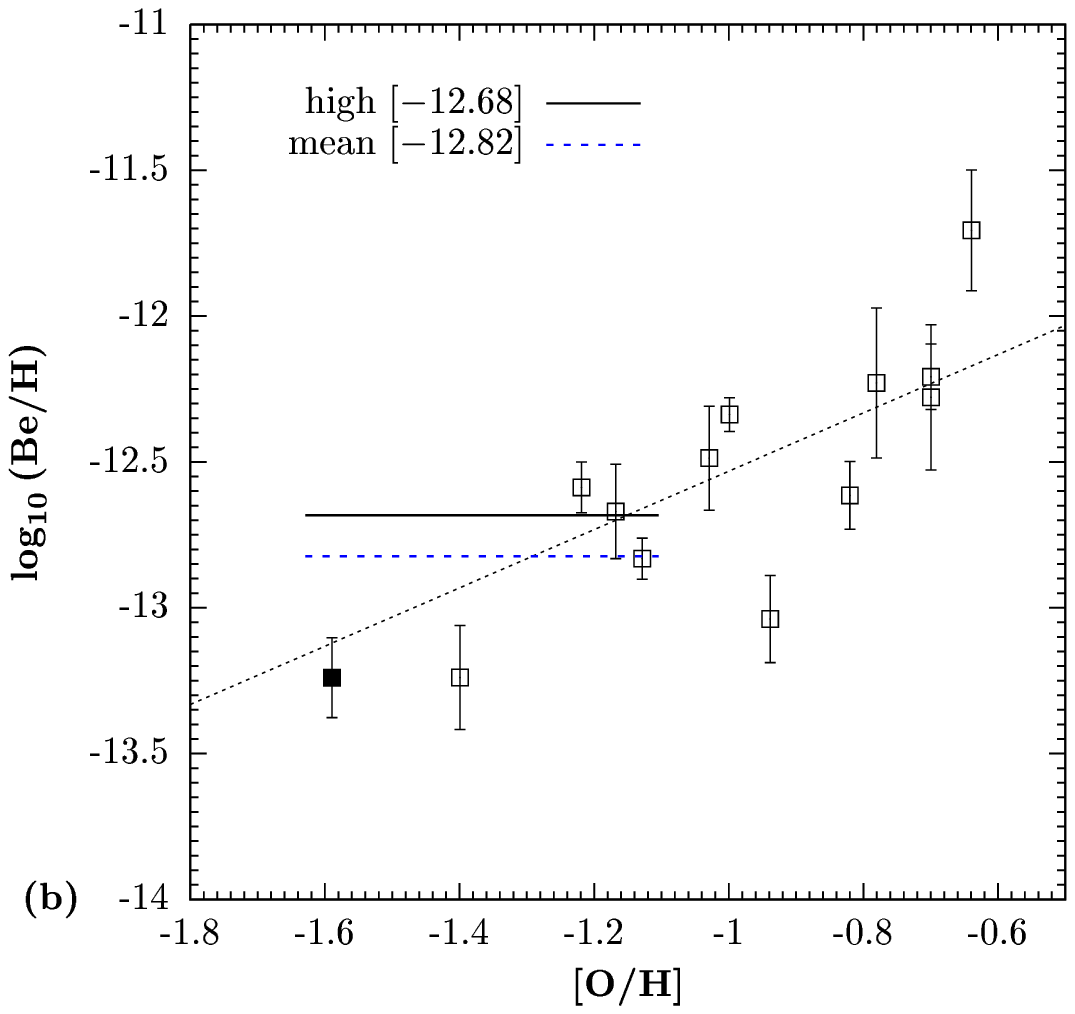}
}

\caption{ \footnotesize Observations of Be in Pop II halo stars. In
  the left panel (a), the data is taken from Fig.~3a of
  Ref.~\cite{Primas:2000gc} and is plotted as a function of
  $[\Fe/\Hyd]$.  The right panel (b) shows the data from Fig.~6b of
  Ref.~\cite{Fields:2004ug} where $[\Ox/\Hyd]$ provides the
  metallicity indicator. The filled dots depict the data points
  associated with the star G~64--12. The solid lines give the inferred
  nominal upper limits on \ben\ from the weighted mean (dashed lines)
  of a sample of stars at lowest metallicity. Also shown in
  Fig.~\ref{Fig:Bedata}b is a fit of a primary scaling of Be; see main
  text.  }
\label{Fig:Bedata} 
\end{figure}

Figure~\ref{Fig:Bedata}a shows the original Be detection in
the star G~64--12%
\footnote{For consistency with the rest of the data points, the 1D LTE
  value has been plotted in Fig.~3a of the original
  reference~\cite{Primas:2000gc}.}
(filled dot) along with a subset of data points taken from Fig.~3a of
Ref.~\cite{Primas:2000gc}.  The data of Fig.~\ref{Fig:Bedata}b are
taken from Fig.~6b of the recent work \cite{Fields:2004ug} which also
uses $[\Ox/\Hyd]$ as a metallicity tracer. The latter paper discusses
the implications of a new temperature scale on the abundances of Li,
Be, and B. In principle, different assumed physical parameters which
characterize the stellar atmosphere may result in large systematic
shifts of the inferred abundances. In this regard, it is important to
note that Be is not overly sensitive to the assumed surface
temperature of the halo dwarfs \cite{Fields:2004ug}. In the following
we thus shall take a pragmatic approach: In both
Fig.~\ref{Fig:Bedata}a and Fig.~\ref{Fig:Bedata}b, we obtain the least
squares weighted mean (dashed lines) for a representative sample of
stars at lowest metallicities. From the variance of the fit, we can
extract a nominal $3\sigma$ upper limit (solid lines) on primordial
\ben. From Fig.~\ref{Fig:Bedata}b, we find%
\begin{equation}
  \label{eq:upper-limit-Be9}
  \log_{10} {\Be/\Hyd}|_{\mathrm{high}} = -12.68 
  \quad\Rightarrow\quad
  \ben/\Hyd \le 2.1\times 10^{-13}\ .
\end{equation}
Conversely, Fig.~\ref{Fig:Bedata}a yields \mbox{$\ben/\Hyd \lesssim
  10^{-13}$} while fitting only the last two data points with
$[\Ox/\Hyd] < -1.3$ in Fig.~\ref{Fig:Bedata}b would give $\ben/\Hyd
\lesssim 1.3\times 10^{-13}$. In our context, those values are less
conservative so that we use (\ref{eq:upper-limit-Be9}) in the
following. 
In Fig.~\ref{Fig:Bedata}b we have additionally fitted for a primordial
component, $\ben/\Hyd|_{\mathrm{p}}$, in combination with a primary
scaling, $\ben/\Hyd=\kappa\, \Ox/\Hyd$. It seems, however, that a
purely primary mechanism with $\kappa\simeq 2.9$ fits the data best
since $\ben/\Hyd|_{\mathrm{p}}$ comes out negligibly small.%
\footnote{For a proper comparison between different assumed surface
  temperature scales and corresponding fits of primary versus
  secondary scaling, see Ref.~\cite{Fields:2004ug}.}
Finally, we are aware that neither of the fitted mean values in
Fig.~\ref{Fig:Bedata} is very good in terms of $\chi^2$.  However, a
firm conjecture of a Be plateau is not the purpose of this work, and
indeed (\ref{eq:upper-limit-Be9}) does provide a sufficiently
conservative limit to work with.

We can now confront the constraint~(\ref{eq:upper-limit-Be9}) with the
CBBN yield of \ben\ obtained by solving the associated Boltzmann
equations. The central input parameter for the catalytic production of
\ben\ and \lisx\ is the abundance of \xm\ at the time of its
recombination with \hef.  Above $10\ \keV$, we can track the resulting
(\hef\xm) abundance by using the Saha-equation since
photo-dissociation proceeds rapidly. Only at $T\simeq 8\ \keV$,
(\hef\xm) starts to build up efficiently, and we couple it into the
full set of Boltzmann equations.  We use a \hef--\xm\ recombination
cross section that is based on the work of
Ref.~\cite{Pospelov:2006sc}.  It takes into account the finite size of
the nucleus and includes $\alpha$-captures into 1S as well as 2S
states. For the cross section of catalyzed \lisx\ production
(\ref{Eq:CBBNLiSix}), we employ the result of a nuclear three-body
calculation~\cite{Hamaguchi:2007mp}.  The path to \ben\ proceeds via
($\beetm$\xm) bound states which are formed by the radiative fusion
$\hef + (\hef\xm) \rightarrow (\beetm\xm)+\gamma$. From there, \ben\
is subsequently produced by resonant neutron capture
$(\beetm\xm)+n\rightarrow\ben+\xm$ \cite{Pospelov:2007js}; see
Appendix~\ref{sec:appendix} for the associated cross sections.  In our
code we can neglect the formation of ($\beetm$\xm) that proceeds via
molecular bound states (\hef$X^-_2$) \cite{Pospelov:2007js}. This
process becomes important only for a combination of large \YX\ and
large \tauX, i.e., a parameter region which is already excluded by
\lisx\ overproduction.  Also note that at the time when (\beet\xm)
form, their photo-dissociation is not important because of the high
binding energy $E_{\mathrm{b}}^{(\beet\xm)} \simeq 1430\
\keV$~\cite{Pospelov:2006sc}. Finally, for $T_9 < 0.2$, the SBBN $n$
abundance can already be tracked well by including the processes
$\Deut + \Deut \to n + \het$, $\Trit + \Deut \to n + \het$, and $\het
+ n \to p + \Trit$ in the reaction network \cite{Mukhanov:2003xs}.
Those cross-sections as well as the one from $p$-induced \lisx\
destruction can be found, {\em e.g.}, in~\cite{Caughlan:1987qf}.  It
is important to note that we assume the SBBN central value for the
deuterium abundance. The early decays of \xm\ may result in an
injection of nucleons into the system. This typically drives the
deuterium abundance upward, resulting in an enhanced number of
neutrons at later times and therefore in an increased output of \ben,
with the general scaling $\benm\sim {\rm const} \times ({\rm D/D_{\rm
    SBBN}})^2$. We choose to disregard this effect, noting its
model-dependent character. We are allowed to do so since its inclusion
can only make the \ben-derived bound on the \xm\ abundance {\em
  stronger}.

We should mention at this point that anyone attempting a precision
calculation of \ben\ within the CBBN framework would have to include
an additional channel related to the {\em early} production of
beryllium as pointed out in \cite{Bird:2007ge}. This channel consists
of \xm\ capture on \bes\ with a subsequent $p$-induced reaction
producing ($^8$B\xm), which then beta decays to the (\beet\xm) bound
state:
\begin{equation}
\label{earlychain}
X^- \to (\besm X^-) \to (^8{\rm B}X^-) \to (\beetm X^-) \to \benm.
\end{equation}
The efficiency of this chain is directly proportional to a rather
small \bes\ abundance. Therefore, the final output of $\ben/\Hyd$ via
(\ref{earlychain}) is never very large, but could reach the level of
$\sim O(10^{-13})$ for large abundances of \xm.  For the purpose of
setting limits on particle physics models, we are allowed to ignore
this early chain (\ref{earlychain}), noting that it is generally
subdominant and also model-dependent.  In particular, the chain
(\ref{earlychain}) depends on the properties of \xm\
\cite{Bird:2007ge}, as well as on the non-thermal processes that can
affect the \bes\ abundance%
\footnote{In addition, as pointed out in recent
  Ref. \cite{Kamimura:2008fx}, the beta decay of $^8$B occurs
  predominantly to the excited states of $^8$Be, which will likely
  result in a break-up of ($^8$Be\xm) and further reduction of the
  efficiency of the chain (\ref{earlychain}).}.

%
%
\begin{figure}[tb]
\centerline{\includegraphics[width=11cm]{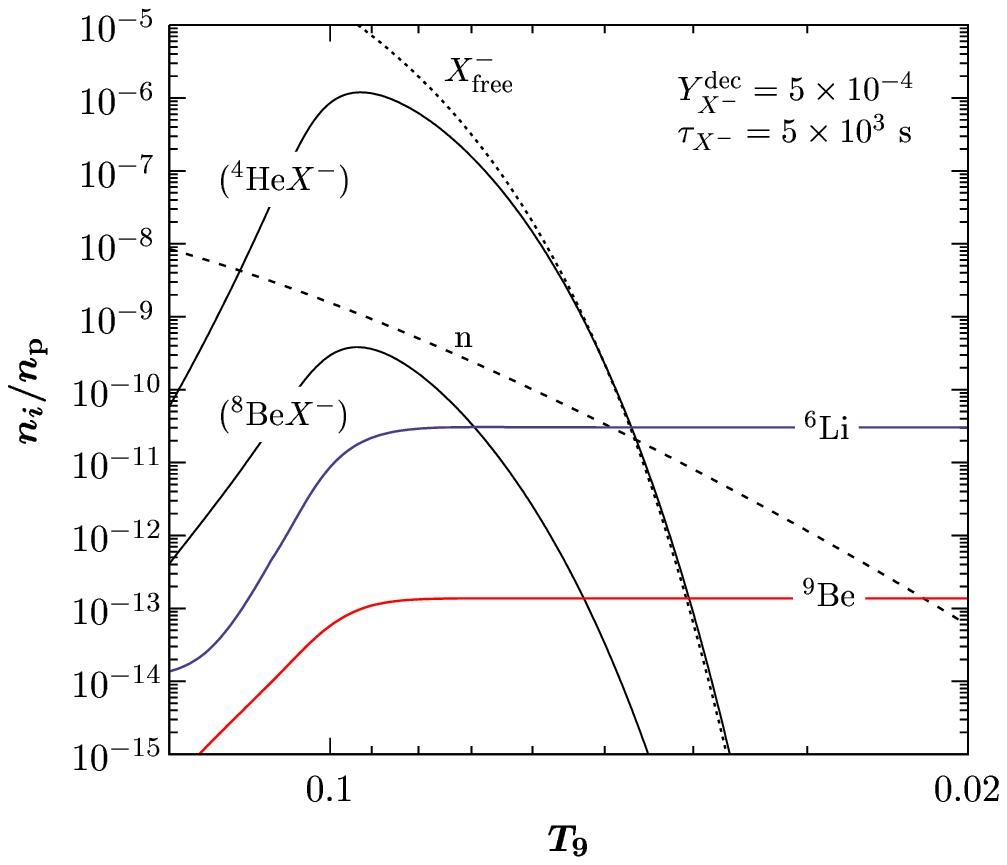}}
\caption{ \footnotesize Evolution of catalyzed \lisx\ and \ben\
  production shown together with the formation of the ``bottle-neck''
  abundances of (\hef\xm) and (\beet\xm) for $\YXdec=5\times 10^{-4}$
  and $\tauX=5\times10^3\ \mathrm{s}$. The dashed line gives the
  neutron abundance while the dotted line shows the abundance of free
  \xm.}
\label{Fig:BeEvol} 
\end{figure}
%
Figure~\ref{Fig:BeEvol} shows the evolution of catalyzed \lisx\ and
\ben\ production from the solution of the corresponding set of
Boltzmann equations below $T=10$ keV.  We parameterize \YX\ by the
\xm\ abundance prior to decay by introducing $\YXdec$, where the
superscript ``dec'' stands for decoupling, and by the \xm\ lifetime
$\tauX$, so that the (total) \xm\ abundance at any moment during BBN
is given by $\YX(t)=\YXdec\times\exp(-t/\tau_X)$. In particular, to
obtain the curves in Fig.~\ref{Fig:BeEvol}, the values $\YXdec =
5\times 10^{-4}$ and $\tauX = 5\times10^3\ \mathrm{s}$ are used. When
the ``bottle-neck'' abundances of (\hef\xm) and (\beet\xm) form, the
catalytic paths (\ref{eq:traf-Li6}) and (\ref{eq:traf-Be9}) to \lisx\
and \ben\ open up, resulting in the asymptotic values $\ben/\Hyd
\simeq 10^{-13}$ and $\lisx/\Hyd \simeq 3\times 10^{-11}$.  The dashed
line shows the neutron abundance and the dotted line the free \xm\
abundance, which is dominated by its exponential decay.  We remark in
passing that residual recombinations of \hef\ with \xm\ lead to the
crossing of the (\hef\xm) and $X^-_{\mathrm{free}}$ lines at late
time.

%
\begin{figure}[tb]
\centerline{\includegraphics[width=12cm]{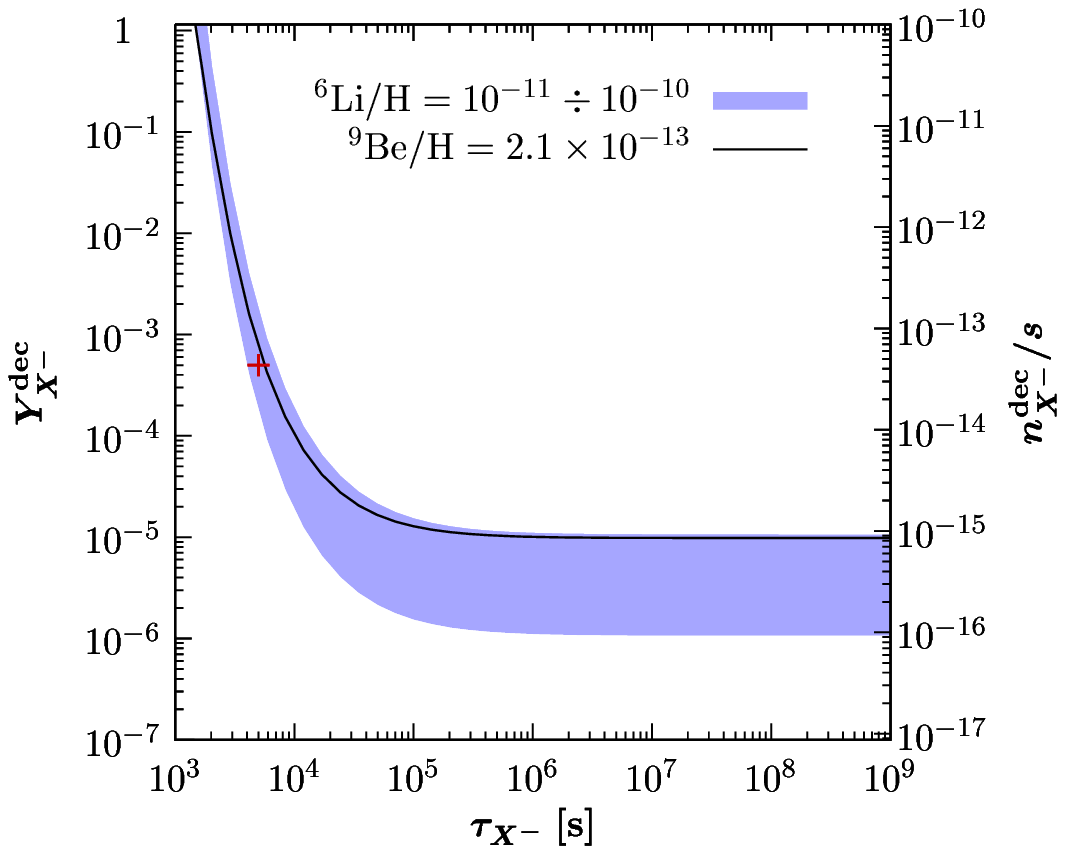}}
\caption{ \footnotesize Contour plot of CBBN abundance yields of
  \lisx\ and \ben\ in the $(\tau_{X^-},Y^{\mathrm{dec}}_{X^-})$ plane.
  The solid line shows the limit (\ref{eq:upper-limit-Be9}). The
  region above this line is excluded by \ben\ overproduction.  The
  lower (upper) boundary of the band corresponds to $\lisx/\Hyd =
  10^{-11}\ (10^{-10})$. The y-axis on the right-hand side indicates
  the \xm\ number density $n^{\mathrm{dec}}_{X^-}$ normalized to the
  entropy density $s$. The cross shows the parameter point considered
  in Fig.~\ref{Fig:BeEvol}. }
\label{Fig:Ytau} 
\end{figure}
%
In Fig.~\ref{Fig:Ytau} we obtain exclusion boundaries from catalyzed
\ben\ and \lisx\ production in the (\tauX,\,\YXdec) parameter space.
For convenience of the reader, the \xm\ number density
$n^{\mathrm{dec}}_{X^-}$ normalized to the entropy density $s$ is
given on the $y$-axis on the right-hand side. Above the solid line,
\ben\ is in excess with respect to (\ref{eq:upper-limit-Be9}) and thus
excluded. The shown band reflects the uncertainties in the
observational determination of \lisx. On the lower border, $\lisx/\Hyd
= 10^{-11}$ is fulfilled while $\lisx/\Hyd = 10^{-10}$ holds on the
upper border of the band. The cross indicates the exemplary parameter
point considered in Fig.~\ref{Fig:BeEvol}. At large lifetimes, the
linear scaling of \lisx\ with \YX\ can easily be seen from the
boundaries of the band.  Note that we find $\ben / \lisx $ in the
interval between $10^{-3}$ and $10^{-2}$, whenever CBBN is efficient,
which confirms the observation already made in
Ref.~\cite{Pospelov:2007js}.

\section{Implications for Supersymmetric Models}
\label{Sec:SUSYImplications}

\begin{figure}[t!]
\centerline{\includegraphics[width=10cm]{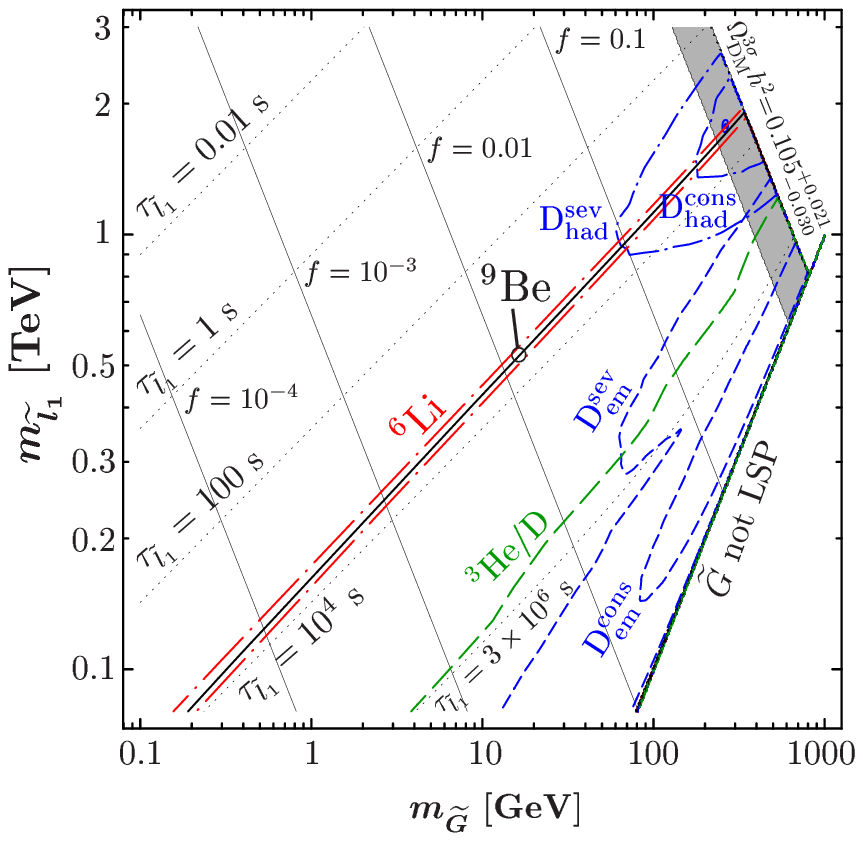}}
\caption{ \footnotesize Cosmological constraints on the masses of the
  gravitino LSP and a charged slepton NLSP for a yield $\Ysldec$ given
  by~(\ref{Eq:Yslepton}). The gray band indicates
  $\Omega_{\gravitino}^{\NTP}\!\!\in\OmegaDM^{3\sigma}$. Above this
  band, $\Omega_{\gravitino} > 0.126$. On the thin solid line labeled
  with $f$ values only $f\,\OmegaDM$ is provided by $\Omegantp$. The
  dotted lines show contours of $\tau_{\slepton}$.  Due to CBBN, the
  region below the solid and the long-dash-dotted (red in the web
  version) lines is disfavored by observationally inferred abundances
  of $^9$Be and $^6$Li, respectively. The effect of electromagnetic
  and hadronic energy injection on primordial D disfavors the regions
  inside the short-dash-dotted (blue in the web version) curves and to
  the right or inside of the short-dashed (blue in the web version)
  curves, respectively. The region below the dashed (green in the web
  version) line is disfavored by the effect of electromagnetic energy
  injection on~$^3\mathrm{He}/\mathrm{D}$. While the constraints from
  hadronic energy injection are obtained for a purely `right-handed'
  $\slepton\simeq\sleptonR$ NLSP, the ones from electromagnetic energy
  injection are valid for the $\stau$ NLSP case with a visible
  electromagnetic energy of $E_{\mathrm{vis}}=\epsilon_{\EM}=0.3
  E_{\tau}$ released in $\stau\to\gravitino\tau$.}
\label{Fig:MstMgrFDS} 
\end{figure}

Let us now address the implications of the results derived above to
SUSY extensions of the Standard Model in which the
gravitino $\gravitino$ is the LSP and a charged slepton $\slepton$ the
NLSP~\cite{Asaka:2000zh,Ellis:2003dn,Fujii:2003nr,Feng:2004mt,Cerdeno:2005eu,Jedamzik:2005dh,Steffen:2006hw,Cyburt:2006uv,Steffen:2006wx,Pradler:2006hh,Pradler:2007is,Kersten:2007ab,Pradler:2007ar,Steffen:2008bt}.
As the spin-3/2 superpartner of the graviton, the gravitino is an
extremely weakly interacting particle with supergravity
couplings~\cite{Cremmer:1982en,Wess:1992cp} that are suppressed by the
(reduced) Planck scale~\cite{Yao:2006px} $\MPl=2.4\times
10^{18}\,\GeV$.
Thereby, the negatively charged $\slepton^-$ can be the long-lived
$X^-$ with a lifetime $\tau_X=\tau_{\slepton}$ governed by the decay
$\slepton\to\gravitino\tau$,%
\footnote{We assume R-parity conservation. For the case of broken
  $\mathrm{R}$-parity, see {\em e.g.}~\cite{Buchmuller:2007ui}.}
\begin{equation}
        \tau_{\slepton} 
        \simeq
        \Gamma^{-1}(\slepton\to\gravitino\lepton)
        = 
        \frac{48 \pi \mgr^2 \MPl^2}{\mslepton^5} 
        \left(1-\frac{\mgr^2}{\mslepton^2}\right)^{-4}
\, .
\label{Eq:SleptonLifetime}
\end{equation}
Indeed, $\tau_{\slepton}\gtrsim 10^4\,\seconds$ occurs in a large
region of natural values of the gravitino mass $\mgr$ and the slepton
mass $\mslepton$, as illustrated by the $\tau_{\slepton}$-contours
(dotted lines) in Fig.~\ref{Fig:MstMgrFDS}.
%

For a standard cosmological history with a post-inflationary reheating
temperature $\TR$ above the decoupling temperature of the $\slepton$
NLSP, $T_{\mathrm{f}}\simlt\mslepton/20$~\cite{Asaka:2000zh}, the
$\slepton$ NLSP freezes out of the primordial plasma as a cold thermal
relic
so that its yield after decoupling $\Ysldec$ is governed by its mass
and its annihilation rate.
Thereby, $\Ysldec$ becomes sensitive to the mass spectrum and the
couplings of the SUSY model.
In this work, we work with a representative yield that is quite
typical for an electrically charged massive thermal
relic~\cite{Asaka:2000zh,Fujii:2003nr,Pradler:2006hh}%
\footnote{For a recent thorough study of the decoupling yield of a
  charged relic, see Ref.~\cite{Berger:2008ti}.}
\begin{equation}
  \Ysldec
  \equiv \frac{n_{\slepton}}{n_{\mathrm B}}
  = 2\,\Ysldecm
  = 0.8 \times 10^{-3}\left(\frac{\mslepton}{100~\GeV}\right)
  ,
\label{Eq:Yslepton}
\end{equation}
where $n_{\slepton}$ denotes the total $\slepton$ number density
assuming an equal number density of positively and negatively charged
$\slepton$'s.
Note that the yield~(\ref{Eq:Yslepton}) is in good agreement with the
curve in Fig.~1 of Ref.~\cite{Asaka:2000zh} that has been derived for
the case of a purely `right-handed' $\stau\simeq\stauR$ NLSP with a
mass that is significantly below the masses of the lighter selectron
and the lighter smuon, $m_{\stau} \ll m_{\sel,\smu}$, and with a
bino-like lightest neutralino, $\neutralino\simeq\Bino$, that has a
mass of $m_{\Bino}=1.1\,m_{\stau}$.
In the case of an approximate slepton mass degeneracy, $m_{\stau}
\lesssim m_{\sel,\smu} \lesssim 1.1\,m_{\stau}$, the $\stau$ NLSP
yield~(\ref{Eq:Yslepton}) can become twice as large due to slepton
coannihilation processes~\cite{Asaka:2000zh,Pradler:2006hh}.
Approaching the $\neutralino$--$\stau$ coannihilation region, $\mneu
\approx \mst$, even larger enhancement factors occur; see {\em
  e.g.}~Fig.~3 in Ref.~\cite{Pradler:2006hh}.
On the other hand, a sizable left--right mixing of the stau NLSP is
associated with an increase of its MSSM couplings and thus with a
reduction of $Y_{\slepton}$.
Moreover, an exceptional reduction of $Y_{\slepton}$ can occur also in
a non-standard thermal history with late-time entropy production after
the decoupling of the $\slepton$ NLSP and before
BBN~\cite{Buchmuller:2006tt,Pradler:2006hh,Hamaguchi:2007mp} or in low
$\TR$ scenarios~\cite{Takayama:2007du}.
Noting that both cases require substantial modifications to the MSSM
field content at or below the weak scale, we disregard such
possibilities and focus in the remainder of this work on the more
generic $Y_{\slepton}$ values described by~(\ref{Eq:Yslepton}).

\begin{figure}[t]
\centerline{\includegraphics[width=10cm]{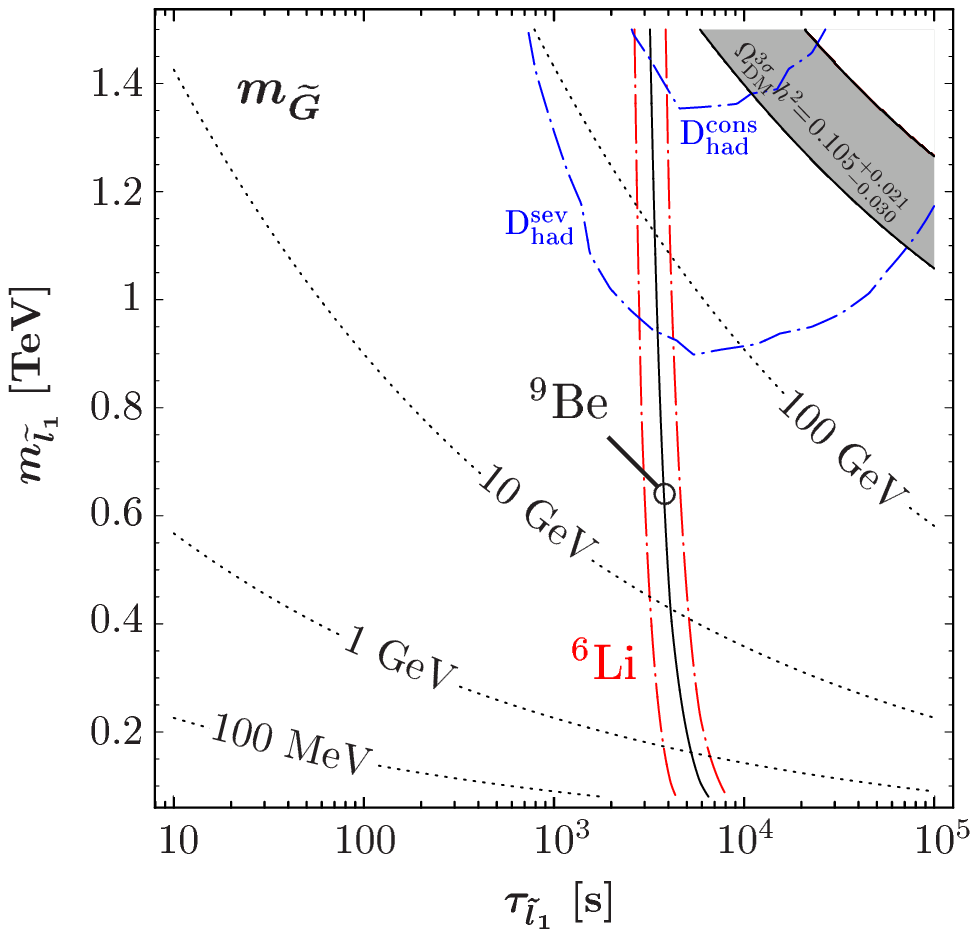}}
\caption{ \footnotesize Contours of $\mgr$ (dotted lines) as a
  function of $\tau_{\slepton}$ and $\mslepton$. Assuming a slepton
  yield $\Ysldec$ given by~(\ref{Eq:Yslepton}), constraints from CBBN
  of $^9$Be and $^6$Li are obtained as shown by the solid line and the
  long-dash-dotted (red in the web version) lines, respectively. For a
  purely `right-handed' $\slepton\simeq\sleptonR$ NLSP with a
  yield~(\ref{Eq:Yslepton}), BBN constraints from effects of hadronic
  energy injection on D are obtained as indicated by the
  short-dash-dotted (blue in the web version) lines.  In the shaded
  region, $\Omegantp h^2$ agrees with~(\ref{Eq:OmegaDM}).  }
\label{Fig:MgrConstraints} 
\end{figure}

Confronting~(\ref{Eq:Yslepton}) with our limits shown in
Fig.~\ref{Fig:Ytau}, we obtain the CBBN constraints shown in
Figs.~\ref{Fig:MstMgrFDS}, \ref{Fig:MgrConstraints},
and~\ref{Fig:TRConstraints}. In each figure, it is the region to the
right of the long-dash-dotted (red in the web version) lines and the
one to the right of the solid line in between those lines that is
disfavored by $\LiHprim\leq 10^{-11}$ and $10^{-10}$ and by the $^9$Be
limit~(\ref{eq:upper-limit-Be9}), respectively.
While the emphasis is on the limits from CBBN of $^9$Be (and $^6$Li),
the following additional cosmological constraints are shown 
for comparison:

\begin{itemize}
  
\item Since each $\slepton$ NLSP decays into one $\gravitino$ LSP,
  these decays lead to a non-thermally produced (NTP) gravitino
  density~\cite{Asaka:2000zh,Feng:2004mt}:
\begin{equation}
        \Omegantp h^2
        = 
        \mgravitino\, \Ysldec\, n_{\mathrm B}(T_0) h^2 / \rho_{\mathrm{c}}
\, ,
\label{Eq:GravitinoDensityNTP}
\end{equation}
where $\rho_c/[n_{\mathrm B}(T_0)h^2]=42.1\,\GeV$~\cite{Yao:2006px}.
This contributes to the relic gravitino density $\Omega_{\gravitino}$
which should not exceed the observationally inferred dark matter
density $\OmegaDM$ and
thus imposes an additional constraint on the model.  Accordingly, we
show shaded regions in Figs.~\ref{Fig:MstMgrFDS},
\ref{Fig:MgrConstraints}, and~\ref{Fig:TRConstraints}, in which the
$\Omega_{\gravitino}^{\NTP} h^2$ values obtained
with~(\ref{Eq:Yslepton}) agree with the nominal $3\sigma$ range of
$\Omega_{\CDM}h^2$ inferred with a restrictive six-parameter
``vanilla'' model from the three year data set of the Wilkinson
Microwave Anisotropy Probe (WMAP) satellite~\cite{WMAPplusX}
\begin{equation}
        \OmegaDM^{3\sigma}h^2=0.105^{+0.021}_{-0.030} 
\label{Eq:OmegaDM}
\end{equation} 
with $h=0.73^{+0.04}_{-0.03}$ denoting the Hubble constant in units of
$100\,\km\,\Mpc^{-1}\seconds^{-1}$.
In each figure, the parameter space above the shaded region is
disfavored by the dark matter constraint $\Omegantp\leq \OmegaDM$.
With any additional contribution to $\OmegaDM$---such as an axion
density or a thermally produced gravitino density $\Omegatp$---this
constraint can become even more restrictive.
In Fig.~\ref{Fig:MstMgrFDS} this is indicated by the thin solid
contours labeled with $f=0.1$, $0.01$, $10^{-3}$, and $10^{-4}$, on
which (\ref{Eq:GravitinoDensityNTP}) obtained with~(\ref{Eq:Yslepton})
satisfies $f\,\Omegantp=0.126$, respectively.

\item In a $\slepton$ NLSP decay, Standard Model particles are emitted
  in addition to the gravitino which can affect the abundances of the
  primordial light elements. While the associated
  hadronic/electromagnetic energy release seems to affect the CBBN
  constraints from $^6$Li and $^9$Be only mildly---as discussed in the
  Introduction---it could alter in a substantial way the primordial
  fractions of D/H and
  $^3$He/D~\cite{Feng:2004mt,Cerdeno:2005eu,Steffen:2006hw,Cyburt:2006uv,Kawasaki:2007xb}.
  The effect of hadronic energy injection on primordial D disfavors
  the regions inside the short-dash-dotted (blue in the web version)
  curves shown in Figs.~\ref{Fig:MstMgrFDS},
  \ref{Fig:MgrConstraints}, and~\ref{Fig:TRConstraints}.%
  \footnote{Additional constraints on hadronic energy release are
    imposed by the primordial abundances of $^4$He, $^3$He/D, $^7$Li,
    and
    $^6$Li/$^7$Li~\cite{Sigl:1995kk,Jedamzik:1999di,Jedamzik:2004er,Kawasaki:2004qu,Jedamzik:2006xz,Cyburt:2006uv}.
    However, in the region allowed by the $^9$Be and $^6$Li
    constraints from bound-state effects, i.e., $\tau_{\st}\lesssim
    10^4~\seconds$, the considered D constraint on hadronic energy
    release is the dominant one as can be seen {\em e.g.} in
    Figs.~38--41 of Ref.~\cite{Kawasaki:2004qu} and in Figs.~6--8 of
    Ref.~\cite{Jedamzik:2006xz}.}
  These curves are obtained from the upper limits on $\Ysldec$ that
  are given in Fig.~11 of Ref.~\cite{Steffen:2006hw} as derived from a
  computation of the 4-body decay of a purely `right-handed'
  $\slepton\simeq\sleptonR$ NLSP into the gravitino, the tau, and a
  quark-antiquark pair~\cite{Steffen:2006hw}.
  Note that these upper limits are based on the severe and
  conservative upper bounds on the released hadronic energy (95\% CL)
  obtained in~\cite{Kawasaki:2004qu} for observationally inferred
  values of the primordial D abundance (see references cited
  in~\cite{Kawasaki:2004qu}):
\begin{eqnarray}
(\mathrm{D}/\mathrm{H})_{\mathrm{mean}} 
&=& (2.78^{+0.44}_{-0.38})\times 10^{-5} 
\quad \Rightarrow \quad \mathrm{severe~constraint},
\label{Eq:Dmean}\\
(\mathrm{D}/\mathrm{H})_{\mathrm{high}} 
&=& (3.98^{+0.59}_{-0.67})\times 10^{-5}
\quad \Rightarrow \quad \mathrm{conservative~constraint}. 
\label{Eq:Dhigh}
\end{eqnarray}
Without trying to give extra credence to a rather
high value of D/H in (\ref{Eq:Dhigh}), following~\cite{Kawasaki:2004qu},
we simply take it as a limiting value for D/H. 

  The regions disfavored by electromagnetic energy injection are shown
  in Fig.~\ref{Fig:MstMgrFDS} only. Here it is the region to the right
  or inside of the short-dashed (blue in the web version) curves and
  the region to the right of the long-dashed (green in the web
  version) line that are disfavored by the primordial abundances of D
  and $^3$He/D, respectively.
  These curves are obtained for the stau NLSP case $\slepton=\stau$,
  i.e., for a `visible' electromagnetic energy of
  $E_{\mathrm{vis}}=\epsilon_{\EM}= 0.3\,E_{\tau}$ of the tau energy
  $E_{\tau}=(\mst^2-\mgr^2+m_{\tau}^2)/2\mst$ released in
  $\stau\to\gravitino\tau$,%
  \footnote{For the selectron NLSP, $\slepton=\sel$, with
    $E_{\mathrm{vis}}=\epsilon_{\EM}=(m_{\sel}^2-\mgr^2+m_{\electron}^2)/2m_{\sel}$
    given by the `full' electron energy released in
    $\sel\to\gravitino\electron$, the bounds become more severe as
    shown explicitly in Fig.~12 of Ref~\cite{Steffen:2006hw}. However,
    a comparison with Fig.~\ref{Fig:Ytau} shown above shows that the
    constraints from electromagnetic energy injection will still be
    significantly less restrictive than the CBBN constraints from $^9$Be
    and $^6$Li.}
  where the $\DsevEM$ and $\HeD$ constraints result from the
  $Y_{\NLSP}$ limits given in Fig.~42 of Ref.~\cite{Kawasaki:2004qu}
  and the $\DconsEM$ constraint from the $Y_{\NLSP}$ limit given in
  Fig.~6 of Ref.~\cite{Cyburt:2002uv}; see also Fig.~9 (lower panel)
  in Ref.~\cite{Steffen:2006hw}. As noted before, the elevated content
  of D leads to the enhancement of CBBN-produced \lisx\ and \ben. For
  example, if non-thermal processes boost the deuterium abundance to
  the level of (\ref{Eq:Dhigh}), it would lead to an enhancement of
  the \lisx\ output by a factor of $\sim 2$, while the corresponding
  enhancement factor in the case of \ben\ is about 4.
  
\end{itemize}

Simple comparison shows  that the $^9$Be constraint  (together
with the one from $^6$Li) provides the most restrictive upper limit on
$\mgr$ for a given $\mslepton$ in the collider-accessible region below
$1~\TeV$.
Indeed, in the most conservative case, $f=1$, the dark matter
constraint $\Omegantp\leq \OmegaDM$ disfavors ($\mgr$,\,$\mslepton$)
combinations associated with $\mgr\gtrsim 200~\GeV$ and
$\mslepton\gtrsim 800~\GeV$, which is a mass range that will be
difficult to probe at the LHC or at the International Linear Collider
(ILC).
Moreover, the `electromagnetic' $\dm_{\EM}$ and $^3$He/D constraints
are always less restrictive than the CBBN constraints from
$^9$Be and $^6$Li.
Only the `hadronic' constraint $\dm_{\HAD}$ can potentially compete with
the CBBN constraints. For instance, this occurs for ($\mgr$,\,$\mslepton$)
combinations associated with $\mgr\gtrsim 60~\GeV$ and a heavy
$\slepton$ NLSP with $\mslepton\gtrsim 900~\GeV$.
Accordingly, the presented CBBN constraints from $^9$Be and $^6$Li are
the most relevant ones in the $\mslepton$ range that will be
accessible at the next generation of particle accelerators.

For $\mslepton$ below $1~\TeV$, the new $^9$Be constraint can also be
considered as the most robust BBN constraint.
Indeed, the difference between the contours labeled with $\Dcons$ and
$\Dsev$ demonstrates that the D constraints are associated with a
significant uncertainty related to the assumed upper limit on the
primordial D/H fraction; cf.~(\ref{Eq:Dmean}) and~(\ref{Eq:Dhigh}).
The uncertainty associated with assumed upper limits on the primordial
$^6$Li/H fraction, which can differ by (even more than) an order of
magnitude, is indicated by the difference between the $\lisxm$
contours obtained for $\LiHprim\leq 10^{-11}$ and $10^{-10}$.
Indeed, since the observational status of $^9$Be is in better shape
than the one of $^6$Li and since $^9$Be is less fragile than $^6$Li
and $^7$Li and thus less affected by stellar mechanisms, the $^9$Be
constraint (which is represented by a single line) can be considered
to be more robust than the $\lisxm$ constraint.

Here we would like to emphasize that the $^9$Be and $^6$Li constraints
are the ones that are the least sensitive to the precise value of
$\Ysldec$ in the region $\Ysldecm\simgt 10^{-4}$. This results from
the fact that the limits are very steep in that region, as can be seen
in Fig.~\ref{Fig:Ytau}.  Indeed, a yield that is twice as large
as~(\ref{Eq:Yslepton}) will affect the position of the $^9$Be and
$^6$Li constraints only very mildly. In contrast, such an enhanced
yield---as encountered, {\em e.g.}, in the case of slepton
coannihilations---leads to significant changes of the dark matter
constraint and the BBN constraints associated with
hadronic/electromagnetic energy injection, as can be seen explicitly
in Fig.~17 of Ref.~\cite{Steffen:2006wx}.
It should also be noted that an elevated slepton yield can lead to an
additional non-thermal output of \lisx\ for $\tau_{\slepton}\gtrsim
\mathrm{few}\times 10^2\ \seconds$. This is because energetic
spallation debris of destroyed \hef\ nuclei from slepton decays can
hit ambient \hef\ and thereby fuse
\lisx~\cite{Jedamzik:2004er,Kawasaki:2004qu,Jedamzik:2006xz}. This mechanism
depends sensitively on the hadronic branching ratio $B_{\mathrm{h}}$
of the 4-body slepton decay into the gravitino, the associated lepton,
and a quark-antiquark pair for which typically $B_{\mathrm{h}}
\lesssim 3\times 10^{-3}$ for $\mslepton \lesssim 2\ \TeV$ (see Fig.~5
of Ref.~\cite{Steffen:2006hw}). Indeed, as discussed in
Ref.~\cite{Jedamzik:2007qk}, for those branching ratios, the effect of
CBBN on \lisx\ is the dominant one in the region which is not already
excluded by the D constraint.  Thus, for $\mslepton \lesssim 1.5\
\TeV$, our obtained limits on \lisx\ overproduction are only
marginally affected by the hadronic energy release of
$\slepton$-decays. However, for larger slepton masses, i.e., for
scenarios of large $\Ysldecm$ in conjunction with
$B_{\mathrm{h}}>10^{-3}$, the hadronic production of \lisx\ becomes
efficient so that only a simultaneous treatment of both effects can
decide on the accurate \lisx\ BBN output. Note that this can make our
presented limits on \lisx\ only stronger. Thus, we are on the
conservative side when neglecting such additional contributions.

Considering the CBBN constraints in Figs.~\ref{Fig:MstMgrFDS},
\ref{Fig:MgrConstraints}, and~\ref{Fig:TRConstraints}, one finds that
the constraints from $^9$Be and $^6$Li are in close proximity of each
other. This coincidence is of course related to the fact that the
ratio of maximally allowed values for \ben\ and \lisx\ is in rough
agreement with the efficiency of producing \ben\ and \lisx\ per each
long-lived negatively charged slepton.
Given the close proximity of these limits, it is tempting to speculate
about a possible CBBN origin of primordial abundances of $\lisxm$ and
$\benm$ at lowest metallicities~\cite{Pospelov:2007js}.  This
possibility applies to the gravitino dark matter scenarios that are
located at or only slightly to the left of these constraints.
It will be interesting to see whether also a solution of the $^7$Li
problem can be found in these scenarios. Indeed, several ways to solve
the $^7$Li problem have been proposed that could be relevant in this
region~\cite{Jedamzik:2004er,Jedamzik:2005dh,Pospelov:2006sc,Cyburt:2006uv,Bird:2007ge,Jedamzik:2007cp,KusakabeplusX}.
However, a definitive answer will require an elaborate treatment of
BBN in which all relevant effects from bound-state formation and
electromagnetic/hadronic energy release are included simultaneously.

Having discussed the generic features of the $^9$Be
constraint and its comparison with other BBN constraints, 
 we now would like to address its implications.
 Our task is facilitated by the $^9$Be constraint being very close to
 the one from $^6$Li, which makes those implications similar to the
 ones of the $^6$Li
 constraint~\cite{Pospelov:2006sc,Cyburt:2006uv,Steffen:2006wx,Pradler:2006hh,Hamaguchi:2007mp,Kawasaki:2007xb,Pradler:2007is,Kersten:2007ab,Pradler:2007ar,Steffen:2008bt}.
 In this respect, Sect.~\ref{Sec:pXcatalysis} becomes important in
 which we show that the possibility of allowed islands in the
 parameter region with large $Y_{\slepton}$/large
 $\tau_{\slepton}$---which was advocated to remain viable in
 Ref.~\cite{Jedamzik:2007cp}---does not exist.
Our present work does thereby reassure the conclusions drawn from the
$^6$Li constraint in a decisive way :
%
\begin{figure}[t]
\centerline{\includegraphics[width=10cm]{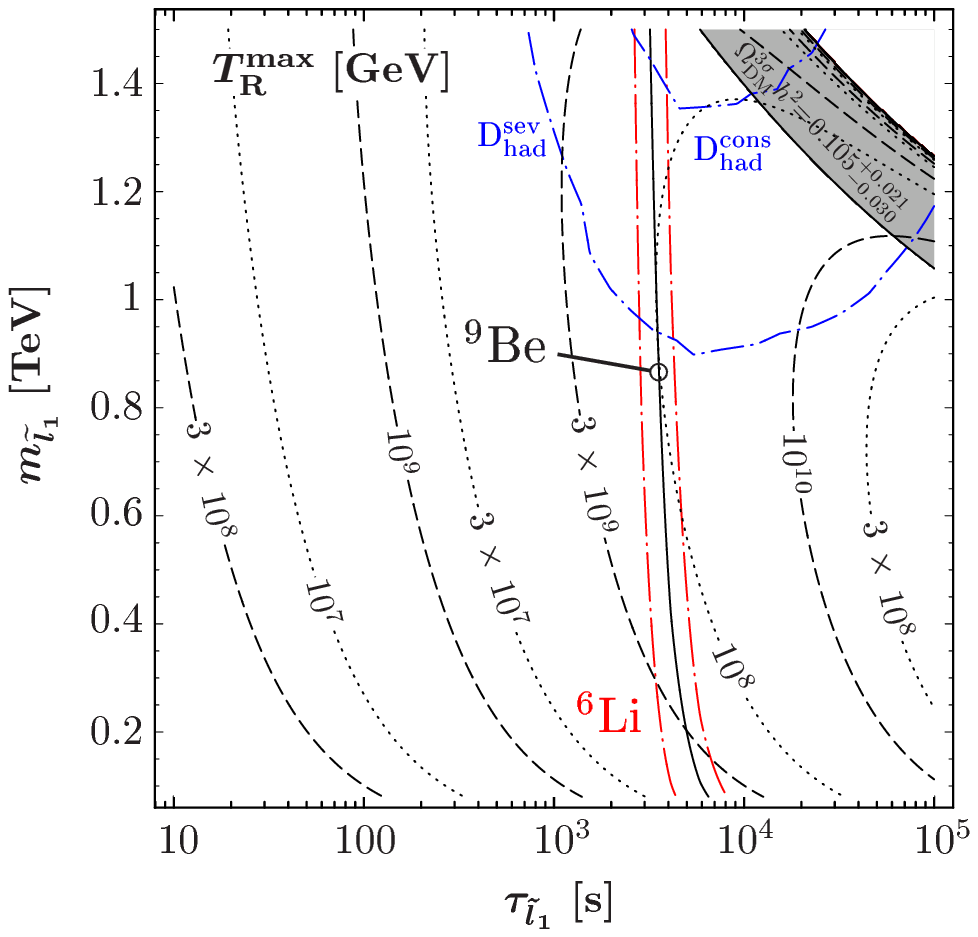}}
\caption{ \footnotesize Contours of $\TRmax$ imposed by $\Omegatp
  h^2+\Omegantp h^2\leq\OmegaDM h^2\leq 0.126$ for $c=1$ (dashed
  lines) and $c=7$ (dotted lines) as a function of $\tau_{\slepton}$
  and $\mslepton$.  The other cosmological constraints are identical
  to the ones shown in Fig.~\ref{Fig:MgrConstraints}.}
\label{Fig:TRConstraints} 
\end{figure}

\begin{enumerate}
  
\item The gravitino mass $\mgr$ is constrained to values well below
  10\% of the slepton NLSP mass $m_{\slepton}$ for $\mslepton
  \simlt\mathcal{O}(1\ \TeV)$. This can be read off conveniently from
  the CBBN constraints shown in Fig.~\ref{Fig:MstMgrFDS} and also from
  Fig.~\ref{Fig:MgrConstraints} in which $\mgr$ contours (dotted
  lines) are given as a function of $\tau_{\slepton}$ and $\mslepton$.
  In particular, $0.1\,{\mslepton}\simlt\mgr< \mslepton$ and thereby
  the kinematical determination of $\mgr$ proposed
  in~\cite{Buchmuller:2004rq} remains cosmologically disfavored at
  least at the next generation of particle accelerators.
  
\item The CBBN constraints disappear for a gravitino mass of
  $\mgr\simlt 200~\MeV$ provided $\mslepton\simgt 80~\GeV$ as
  supported by the non-observation of long-lived charged sleptons at
  the Large Electron Positron Collider (LEP)~\cite{Yao:2006px}. This
  can be seen in Fig.~\ref{Fig:MstMgrFDS}.  Accordingly, for
  gauge-mediated SUSY breaking leading to small values of $\mgr$, the
  CBBN constraint can be irrelevant.
  However, for $\mgr\simgt 10~\GeV$, as obtained in gravity-mediated
  SUSY breaking, the CBBN constraints impose a lower limit of
  $\mslepton>400~\GeV$ as can be seen in Figs.~\ref{Fig:MstMgrFDS}
  and~\ref{Fig:MgrConstraints}.
  With CMSSM relations between the masses of the superparticles, this
  translates into a lower limit on the gluino mass
  $m_{\gluino}>2.5~\TeV$~\cite{Pradler:2007is,Kersten:2007ab,Pradler:2007ar}.
  Thereby, the $^9$Be constraint also points within the CMSSM to a
  cosmologically favored mass range that will be difficult to probe at
  the LHC.

\item The $^9$Be constraint imposes an upper limit on the lifetime
  $\tau_{\slepton}$ that ranges between $6\times 10^3\,\seconds$ at
  $\mslepton=100~\GeV$ and $3\times 10^3\,\seconds$ at
  $\mslepton=1.5~\TeV$ for a charged slepton NLSP with $\Ysldec$
  described by~(\ref{Eq:Yslepton}), as can be seen in
  Fig.~\ref{Fig:MgrConstraints}.
  Indeed, when the value of $\Ysldec$ increases by a factor of 15, the
  $\tau_{\slepton}$ limit decreases by only a factor of two.  This
  mild $\mslepton$ dependence of the $\tau_{\slepton}$ limit reflects
  the fact that the $^9$Be constraint is quite insensitive to the
  precise value of $Y_{\slepton}$.
  While the $\mgr$ contours and the D constraints shown in
  Fig.~\ref{Fig:MgrConstraints} are specific to the gravitino LSP
  scenario with unbroken R-parity, the CBBN limits on
  $\tau_{\slepton}$ shown in Fig.~\ref{Fig:MgrConstraints} apply to
  any scenario with a long-lived charged slepton described by the
  yield~(\ref{Eq:Yslepton}) and can thus be relevant for axino LSP
  scenarios~\cite{Covi:1999ty+X,Brandenburg:2005he,Steffen:2007sp}
  and scenarios with R-parity violation~\cite{Buchmuller:2007ui} as
  well.
  
\item Gravitino dark matter can originate not only from NLSP decays
  but also from thermal scattering of particles in the hot primordial
  plasma. Thereby, the relic gravitino density $\Omega_{\gravitino}$
  receives an additional contribution $\Omegatp$ that depends
  basically linearly on the reheating temperature $\TR$ after
  inflation~\cite{Bolz:2000fu,Steffen:2006hw,Pradler:2006qh,Rychkov:2007uq}.
  In turn, the dark matter constraint,
  $\Omegatp+\Omegantp\leq\OmegaDM$, can be translated into a
  conservative upper limit~\cite{Steffen:2008bt}
\begin{equation}
  \TR 
  \leq 
  \frac{2.37\times 10^9~\GeV}{c^{2}} 
  \left(\frac{\OmegaDM h^2-\Omegantp h^2}{0.1}\right)
  \left(\frac{\tau_{\slepton}}{10^4~\seconds}\right)^{\!\!\frac{1}{2}}\!\!
  \left(\frac{\mslepton}{100~\GeV}\right)^{\!\!\frac{1}{2}}\!\!
  \equiv
  \TR^{\max}
  \, ,
\label{Eq:TR_max}
\end{equation}
which depends on the ratio of the gluino mass $m_{\gluino}$ and the
$\slepton$ NLSP mass, $c\equiv m_{\gluino}/\mslepton>1$, at the weak
scale.
In Fig.~\ref{Fig:TRConstraints}, contours of $\TRmax$ obtained with
$\OmegaDM h^2\leq 0.126$ and $\Omegantp$ given by the
yield~(\ref{Eq:Yslepton}) are shown for the limiting case $c=1$
(dashed lines) and for the case $c=7$ (dotted lines), which is typical
for universal soft SUSY breaking parameters at the scale of grand
unification.
Thus, also with the $^9$Be constraint and with $\Omegantp$ included,
$c=1$ ($c=7$) is found to be associated with a $\TRmax$ value of
$3\times 10^9\,\GeV$ ($10^8\,\GeV$) in the cosmologically favored
region.  Indeed, the $\TR$ constraints for $c=7$ are consistent with
earlier findings within the CMSSM and other constrained
scenarios~\cite{Pradler:2006hh,Pradler:2007is,Choi:2007rh,Kersten:2007ab,Pradler:2007ar}.
Since the $^9$Be constraint imposes $\tau_{\slepton}\simlt 6\times
10^3\,\seconds$ as discussed above, associated upper limits on the
mass ratio $c$ for a given lower limit on the reheating
temperature---such as $\TR> 10^9\,\GeV$ required by successful thermal
leptogenesis with hierarchical heavy right-handed Majorana
neutrinos---can be inferred from Fig.~3 of Ref.~\cite{Steffen:2008bt}.

\end{enumerate}

Let us conclude this section by considering prospects at future
colliders. If the gravitino LSP scenario with a not too heavy charged
slepton NLSP is realized in nature, the production and analysis of the
(quasi-) stable charged sleptons will be a realistic option. Thereby,
collider measurements of $\mslepton$ will become
available~\cite{Ambrosanio:2000ik,Ellis:2006vu}.  Moreover, with an
experimental reconstruction of at least some part of the SUSY model,
one will be able to calculate $\Ysldec$ reliably for a standard
thermal history with $\TR$ above the decoupling temperature
$T_{\mathrm{f}}$ of the slepton NLSP.  Confronting the obtained
$\Ysldecm$ with the CBBN constraints from $^9$Be and $^6$Li shown in
Fig.~\ref{Fig:Ytau} can then provide an upper limit on the lifetime
$\tau_{\slepton}$ of the long-lived slepton. Assuming the gravitino
LSP scenario, this $\tau_{\slepton}$ constraint together with the
measured $\mslepton$ will imply an upper limit on the gravitino mass
$\mgr$ as can be seen in Figs.~\ref{Fig:MstMgrFDS}
and~\ref{Fig:MgrConstraints}. In addition, if $\tau_{\slepton}$ can be
measured, {\em e.g.}, by analyzing $\slepton$ decays in a collider
detector~\cite{Ambrosanio:2000ik,Martyn:2006as} or in some additional
stopper
material~\cite{Hamaguchi:2004df,Feng:2004yi,Brandenburg:2005he,Hamaguchi:2006vu},
one should be able to compare the experimentally determined
combination ($\tau_{\slepton}$,\,$\mslepton$) with the CBBN
constraints shown in Fig.~\ref{Fig:MgrConstraints} (without any
assumption on the gravitino LSP scenario). While a finding of
($\tau_{\slepton}$,\,$\mslepton$) in the region disfavored by CBBN
could point to a non-standard cosmological history with late-time
entropy
production~\cite{Buchmuller:2006tt,Pradler:2006hh,Hamaguchi:2007mp} or
to a low reheating temperature~\cite{Takayama:2007du}, it would be
most remarkable to find ($\tau_{\slepton}$,\,$\mslepton$) in close
vicinity of the CBBN constraints. Notwithstanding a rather large
number of ``if''s compounded in the previous sentences, the collider
measurements could provide in this way an independent test for a
hypothesis of the CBBN origin of $^6$Li and $^9$Be at lowest
metallicities.
Complementary to that it will be exciting to see new analyses of
$^9$Be and $^6$Li data from future astrophysical observations.

\section{Conclusions}
\label{Sec:Conclusions}

From observations of beryllium in Population II halo stars at very low
metallicities, we have extracted a nominal upper limit on primordial
beryllium of $\ben/\Hyd \le 2.1\times 10^{-13}$.  This limit allows
one to set interesting constraints on models in which the primordial
$A=8$ divide is bridged by catalytic effects.  Considering the
primordial catalysis of $^9$Be via bound-state effects of a negatively
charged massive relic $X^-$~\cite{Pospelov:2007js}, we have derived
\tauX-dependent upper limits on the $X^-$ yield prior to decay,
\YXdec.
For a typical relic abundance $\YXdec\simgt 3 \times 10^{-4}$
($10^{-4}$), we find that this $\ben$ limit translates into an upper
limit on the $X^-$ lifetime of $\tauX\simlt 6\times 10^{3}\,\seconds$
($10^{4}\,\seconds$), which is quite comparable with the $\tauX$
limit inferred from the primordial catalysis of $^6$Li.
Moreover, in the region where CBBN is efficient, we confirm that the
ratio of the synthesized elements of $\ben/\lisx$ lies in the range
$10^{-3}-10^{-2}$~\cite{Pospelov:2007js}, which provides perhaps the
most model-independent prediction in the whole CBBN paradigm.

We have clarified that the presence of ($p$\xm) bound states cannot
relax the $\YXdec$ limits at long lifetimes $\tauX$ in any substantial
way. 
Indeed, we have shown explicitly that late-time effects of ($p$\xm)
bound states can affect the lithium and beryllium abundances
synthesized at $T\simeq8\ \keV$ by not more than 10\%.
Any substantial formation of ($p$\xm) at $T\simeq 0.7\ \keV$ is
immediately intercepted by the very efficient charge exchange reaction
of ($p$\xm) with \hef. This comes as no surprise given the large size
of the ($p$\xm) system $\sim 30~\fm$ and the fact that the proton
deconfinement probability approaches unity already for a \hef--\xm\
distance of $\sim 95~\fm$.  In particular, we find that the fractional
density of protons in bound states does not exceed the level of $\sim
10^{-6}$ for $\YX\lesssim Y_{\hef}$.  Correspondingly, even with a
($p$\xm)--induced \lisx-destruction-cross section as large as the
unitarity limit, at most a few percent of the synthesized \lisx\ could
be destroyed. By the same argument, the \ben\ yield also remains
unaffected by late-time catalysis.
Thus, we find that the possibility of allowed islands in the parameter
region with typical $\YXdec$ and large $\tauX$---which was advocated
in Ref.~\cite{Jedamzik:2007cp}---does not exist.
 
Applying the $\tauX$-dependent upper limits on $\YXdec$ derived from
the primordial catalysis of \ben, we have analyzed the new \ben\
constraint in SUSY models in which the gravitino is the LSP and a
long-lived charged slepton the NLSP, $\slepton=X^-$.
For typical values of the slepton NLSP yield after decoupling, the 
$^9$Be constraint obtained in this paper is found in close vicinity to the constraint from
the primordial catalysis of $^6$Li.
Accordingly, the implications of the $^9$Be constraint for SUSY models
do not differ much from the case of $^6$Li.  The important advantage
of \ben-derived constraints is due to the fact that \ben\ is firmly
detected in a significant number of stars at low metallicities, while
the status of \lisx\ observations remains somewhat questionable.
Another great virtue of \ben-derived constraints is due to the
expectations that an \ben\ abundance would remain less affected
compared to the one of \lisx\ by any stellar mechanism that might have
caused the depletion in \lisv.  Therefore, one would not expect any
serious depletion factors between the (hypothetical) primordial
fraction of \ben\ and the actual observationally determined abundances
of \ben.

One could question the calculational status of the CBBN chain
(\ref{eq:traf-Be9}) leading to \ben. The rates for the first two
reactions, resulting in (\hef\xm) and (\beet\xm), are determined by
electromagnetic interactions and thus are not associated with large
nuclear uncertainties.  The final step on the way to \ben\ is the
neutron capture by the (\beet\xm) bound state. It is dominated by a
resonant transition, and the associated rate is believed to have a
factor of a few uncertainty \cite{Pospelov:2007js}.  However, given
the wealth of existing experimental data on the $^9$Be resonances, and
rapid progress in nuclear calculations of few nucleon systems, one
could hope that reasonably precise calculations of the catalytic rates
may become available.

The main shortcoming of our analysis is that only the catalytic
effects below 10 keV are taken into account. In a generic framework of
a hypothetical \xm\ particle, this is fully justified.  However, in
the specific SUSY model with gravitino LSP/slepton NLSP, one could go
one step further and by combining the catalytic effects with the
energy injection effects. Fortunately for us, the question of \ben\
synthesis is somewhat decoupled from the question of energy injection.
The connection is mainly due to the modification of the deuterium
abundance: For example, hadronic energy input at $T\sim 30$ keV would
lead to a larger deuterium abundance, which in turn would enhance the
neutron abundance resulting in a larger abundance of \ben. Thus, the
energy injection would lead to abundances of \ben\ that are somewhat
{\em larger} than the ones that we determined by using the standard
input for the deuterium abundance.  Since the main idea of our paper
is to derive a conservative limit on SUSY models from \ben, we can
stay on the conservative side and disregard this modification, which
would make the \ben\ limit only stronger.

In comparison to other cosmological constraints on gravitino LSP
scenarios with a long-lived charged slepton NLSP, we find that the
$^9$Be constraint (together with the one from $^6$Li) is the most
relevant one in the collider-accessible region of slepton masses below
1~TeV. Indeed, if a SUSY scenario with a long-lived $\slepton$ is
realized in nature, one might be able to determine the combination
($\tau_{\slepton}$,\,$\mslepton$) at collider experiments.  Assuming
standard cosmological history, one might also ``invert'' the collider
data and infer $\Ysldec$. Independently of the assumption of the
gravitino LSP, these quantities can then be confronted with
constraints on the ($\tau_{\slepton}$, $\Ysldec$) parameter space
imposed by the primordial catalysis of $^6$Li and $^9$Be.  Thereby,
the CBBN constraints can be considered as predictions that could be
tested in upcoming high-energy experiments.
It will be most remarkable if collider measurements point to a
($\tau_{\slepton}$,\,$\mslepton$) combination in the vicinity of the
CBBN constraints. Indeed, this could provide an experimental hint for
the primordial catalysis being the origin of existing abundances of
both $^9$Be and $^6$Li at lowest metallicities.  However tenuous the
BBN--LHC connection may seem at the moment, we expect a lot more
clarity brought to this issue in the coming years.

{\bf Note added} -- After the submission of this paper, a dedicated
nuclear physics study of some CBBN reactions has appeared,
Ref.~\cite{Kamimura:2008fx}. It supports the conclusion of this paper
about the large rate for the charge exchange reactions that remove
$(pX^-)$. At the same time, this work finds non-negligible shifts,
$O(100~{\rm keV})$, of the resonant energy levels employed in the
\ben\ production chain.  This may affect the overall efficiency of
\ben\ production, and further investigations of the nuclear rates are
needed.

{\bf Acknowledgments} -- Research at the Perimeter Institute is supported
in part by the Government of Canada through NSERC and by the Province
of Ontario through MEDT. 
M.P.\ would like to thank C. Bird for an independent calculation of
the cross sections for the charge exchange reactions.
J.P.\ is grateful to R.~Lang and Y.Y.Y.~Wong for helpful exchange on
data fitting and to F.~Hahn-Woernle for discussions on the Boltzmann
code.
The authors acknowledge useful conversations with all participants of
the ``BBN and Particle Physics'' workshop held at Perimeter in May of
2008.  The research of J.P.\ and F.D.S.\ is supported in part by the
Cluster of Excellence `Origin and Structure of the Universe.' Both
would like to acknowledge the hospitality of the Perimeter Institute
where part of this work was completed.

\begin{appendix}

\begin{section}{CBBN reaction rates below 10 keV}
\label{sec:appendix}

In the following, we collect the key reaction rates, $ N_{\mathrm{A}}
\langle \sigma v \rangle$, used in the numerical solutions of the
Boltzmann equations. They are given in units of 
$\mathrm{cm}^3\mathrm{s}^{-1}\mathrm{mol}^{-1}$ and $T_9 =
T/10^9~\mathrm{K}$.
\begin{itemize}
\item Recombination and photo-dissociation of \xm:
\begin{align*}
\hef + \xm \to (\hef\xm) + \gamma :\hspace{0.5cm}  & 
7900\; T_9^{-1/2} \\[0.1cm]
(\hef\xm) + \gamma_{\mathrm{bg}} \to \hef + \xm :\hspace{0.5cm}  & 
1.85\times 10^{10}\; T_9^{-2}\; \exp{(-4.03/T_9)} \\[0.1cm]
p + \xm \to (p\xm) + \gamma :\hspace{0.5cm}  & 
3980\; T_9^{-1/2} \\[0.1cm]
(p\xm) + \gamma_{\mathrm{bg}} \to p + \xm :\hspace{0.5cm}  & 
1.18\times 10^{9}\; T_9^{-2}\; \exp{(-0.29/T_9)} 
\end{align*}
\item Charge exchange reactions:
\begin{align*}
(p\xm) + \hef \to (\hef\xm) + p  :\hspace{0.5cm}  & 
3.9\times 10^{10}\; T_9^{1/2}  \\[0.1cm]
(p\xm) + \lisx \to (\lisx\xm) + p  :\hspace{0.5cm}  & 
6.45\times 10^{10}\; T_9^{1/2}  \\[0.1cm]
(p\xm) + (\lisx\xm) \to (\lisx\xm_2) + p  :\hspace{0.5cm}  & 
3.37\times 10^{9}\; T_9^{1/2} \; (1\ \TeV/m_{\xm})^{1/2} \\[0.1cm]
(p\xm) + (\lisx\xm_2) \to (\lisx\xm_3) + p  :\hspace{0.5cm}  & 
5.25\times 10^{8}\; T_9^{1/2} \; (1\ \TeV/m_{\xm})^{1/2} 
\end{align*}
\item \lisx\ and \ben\ catalysis (from \cite{Hamaguchi:2007mp} and
  \cite{Pospelov:2007js}, respectively):
\begin{align*}
  (\hef\xm) + \Deut \to \lisx + \xm  :\hspace{0.5cm}  & 
  2.37\times 10^8\; (1-0.34\, T_9)\, T_9^{-2/3} \exp{(-5.33\, T_9^{-1/3})}  \\[0.1cm]
  \hef + (\hef\xm) \to (\beet\xm) + \gamma :\hspace{0.5cm} &
  10^{5}\; T_9^{-3/2}\; [\ 0.95 \exp{(-1.02/T_9)} \\
  & \hphantom{10^{5}\; T_9^{-3/2}\; \qquad}  + 0.66 \exp{(-1.32/T_9)}\ ]  \\[0.1cm]
(\beet\xm) + n \to \ben + \xm  :\hspace{0.5cm}  & 2\times 10^{9}
\end{align*}
\end{itemize}

\end{section}

\end{appendix}

\end{document}